\documentstyle[12pt]{article}
\let\useblackboard=\iftrue
\useblackboard
\font\blackboard=msbm10 scaled \magstep1
\font\blackboards=msbm7
\font\blackboardss=msbm5
\newfam\black
\textfont\black=\blackboard
\scriptfont\black=\blackboards
\scriptscriptfont\black=\blackboardss

\else

\fi

\newcommand{\ka}{\kappa}
\newcommand{\w}{\omega}
\newcommand{\te}{\tilde {e}}
\newcommand{\ot}{\tilde {o}}
\newcommand{\MM}{{\cal M}}
\newcommand{\td}{\tilde}
\newcommand{\pe}{\perp}
\newcommand{\pa}{\parallel}
\newcommand{\be}{\begin{equation}}
\newcommand{\ee}{\end{equation}}
\newcommand{\ben}{\begin{eqnarray}\displaystyle}
\newcommand{\een}{\end{eqnarray}}
\newcommand{\refb}[1]{(\ref{#1})}
\newcommand{\sectiono}[1]{\section{#1}\setcounter{equation}{0}}

\begin{document}
{}~ \hfill\vbox{\hbox{hep-th/yymmnn}\hbox{}}\break

\vskip 1cm

\centerline{\large \bf Moyal Representation of 
the String Field Star Product}
\vspace*{2.0ex}
{\centerline{\large \bf
in the Presence of a B-background}

\vspace*{7.0ex}

\centerline{
Yong-Shi Wu$$\footnote{email: 
{\tt wu@physics.utah.edu}} 
and Ting-Liang Zhuang$$\footnote{email:
{\tt zhuang@physics.utah.edu}}}

\vspace*{3.0ex}

\centerline{\it Department 
of Physics, University of Utah,
Salt Lake City, UT 84112}
\vspace*{2.8ex}

\vspace*{5.0ex}

\centerline{\bf Abstract}
\bigskip

In this paper we show that in the presence of an 
anti-symmetric tensor $B$-background, Witten's 
star algebra for open string fields persists to 
possess the structure of a direct product of 
commuting Moyal pairs. The interplay between the 
noncommutativity due to three-string overlap and 
that due to the $B$-background is our main concern. 
In each pair of noncommutative directions parallel 
to the $B$-background, the Moyal pairs mix string 
modes in the two directions and are labeled, in 
addition to a continuous parameter, by {\it two} 
discrete values as well. However, the Moyal 
parameters are $B$-dependent only for discrete 
pairs. We have also demonstrated the large-$B$ 
contraction of the star algebra, with one of 
the discrete Moyal pairs dropping out while 
the other giving rise to the center-of-mass 
noncommutative function algebra.

\vfill 
\eject

\baselineskip=16pt

\tableofcontents

\sectiono{Introduction}\label{s0}

Coordinate noncommutativity, which generates 
noncommutativity of functions, lies at the 
heart of noncommutative geometry. Physically,
this implies non-locality of interactions.

In string theory there are two kinds of 
noncommutativity. The first one is that 
proposed by Witten \cite{EW} in his 
cubic Open String Field Theory (OSFT). 
Witten's star product of string fields 
describes a noncommutative geometry for 
the state space of string theory. The 
noncommutativity is due to the gluing 
procedure that defines the star product, in 
which the right-half of the first string is 
glued with the left-half of the second string, 
with the resulting (third) string composed of 
the left-half of the first and the right-half 
of the second. This three-string vertex 
involves $\delta$-functional overlap interactions. 
It is the {\it non-local nature} of the latter 
that gives rise to noncommutativity of Witten's 
star product. (Though $\delta$-functional overlap 
interactions may look quite formal, the three-string 
vertex has a precise oscillator formulation, 
developed in refs. \cite{GJ,CS}.) The second 
kind of noncommutativity is that of the end 
points of an open string in the presence of 
an anti-symmetric $B$-background 
\cite{doug-hull,chuho,sch,sw}, which can be viewed 
as a descendant of the noncommutativity in 
Matrix theory \cite{cds,howu}. In open string 
theory, an anti-symmetric tensor $B$-background 
can be traded off with a gauge field acting on 
the string ends. In an appropriate double 
scaling limit that decouples effects of 
gravity \cite{sw}, an open string behaves like 
a dipole in the lowest Landau level of a magnetic 
field \cite{suss}. It is the non-locality due 
to the finite size of a dipole in the lowest 
Landau level that makes the resulting theory 
in the decoupling limit a noncommutative gauge 
theory. Mathematically this gauge theory can 
be obtained \cite{sw,cr,li} by deforming the 
ordinary product of functions into the Moyal 
star product, a procedure familiar in the 
deformation quantization scheme for quantum 
mechanics. 
  
One may wonder whether there is a close 
relation or an interplay between these two 
kinds of noncommutativity in string theory.
Recently important progress has been made 
on better understanding of this issue. First, 
Bars \cite{Bars} has succeeded in identifying 
Witten's star product of string fields with 
Moyal's star product between canonical pairs 
of appropriate linear combination of string 
modes. Secondly, Rastelli, Sen and Zwiebach 
\cite{RSZ} (RSZ) have successfully solved the 
spectrum, both eigenvalues and eigenvectors, 
of the Neumann coefficient matrices that 
appear in the oscillator formulation of 
three-string vertex in the zero-momentum 
matter sector. This breakthrough makes 
it possible to diagonalize the three-string 
vertex in oscillator formalism. Indeed, soon 
after ref.\cite{RSZ}, a new basis was found
by Douglas, Liu, Moore and Zwiebach \cite{DLMZ} 
in the single string Hilbert space that 
diagonalizes Witten's star product into a 
continuous tensor product of mutually 
commuting two-dimensional Moyal star products. 
The noncommutativity parameter $\theta(\ka)$ 
for each of the Moyal products is given as 
a function of the eigenvalue $\lambda(\ka)$,
where $\ka \in [0,\infty)$. The Moyal 
coordinates $x(\ka)$ and $y(\ka)$ for a 
single string are constructed with the 
twist-even and twist-odd RSZ eigenvectors 
respectively. Shortly in ref. \cite{Belov},
the diagonal representation of the open 
string star product was generalized to 
include the zero modes (center-of-mass 
coordinates) as well. 

It would be of great interests to see how 
the noncommutativity due to a $B$-background 
in target spacetime would intertwine with 
that of the string field star product to
form a bigger structure, Moyal or not. The 
string field star product has been studied 
in e.g. refs. \cite{Sugino,KT,BMS}). Also the 
spectrum (eigenvalues and eigenvectors) 
of the Neumann coefficient matrices in a 
$B$-background in two spatial directions
has been solved recently in refs. 
\cite{FHM,CL}. Based on these results, 
in this paper we study the full structure 
of the open string field algebra, with both 
zero modes and a background $B$-field included. 
It is shown that in the presence of an 
anti-symmetric tensor $B$-background, Witten's 
star algebra for open string fields persists to 
possess the structure of a direct product of 
commuting Moyal pairs. Assume that the 
$B$-background is block-diagonal with $2\times 2$ 
blocks. Then in each pair of noncommutative 
directions along which the B-field has non-zero 
components, the Moyal pairs mix string modes in 
the two directions and are labeled, in addition 
to a continuous parameter, by {\it two} discrete 
values as well. However, the presence of a 
$B$-field affects only the Moyal parameters for 
the pairs with discrete label, while the Moyal 
parameters for the pairs with continuous label, 
even in the noncommutative directions, remain 
unaffected at all. Moreover, we have also 
demonstrated the large-$B$ contraction of the 
star algebra, noticing a singular behavior of 
one of the discrete Moyal pairs: It drops out 
the contraction, while the other pair gives 
rise to the center-of-mass noncommutative 
function algebra. Previously the large-$B$ 
limit was discussed in the literature 
\cite{EW2,Schnabl} in terms of the vertex 
operators or string oscillators; here we 
examine the contraction directly in terms 
of the Moyal modes in the three-string 
vertex. The singular behavior we found for
the large-$B$ limit may require some caution
be taken when an argument with the help of
the large-$B$ contraction is made. (For recent 
papers on other aspects of the continuous 
Moyal product in OSFT, see refs. 
\cite{AG,BM,Erler,CHL,BK}.)

The paper is organized as follows. In Section 2 
and Section 3, using the eigenvectors of Neumann 
matrices, we introduce two sets of new oscillators, 
and rewrite the three-string vertex, first with 
zero modes and then in a $B$-background, explicitly 
in terms of them. In Section 4 we identify the 
Moyal structure for the open string star product, 
and present the explicit expression for 
noncommutativity parameters as well as for the 
Moyal coordinates. In Section 5 we demonstrate 
the large-$B$ contraction of the star algebra. 
Finally a summary of our results and discussions 
are given in Section 6.

\sectiono{Three-string vertex with zero modes} \label{s1}

Witten's star product of string fields is defined 
in terms of a (non-local) half-string overlap. In 
oscillator representation \cite{GJ}, this product 
can be formulated using a three-string vertex 
involving a quadratic form of string modes, whose 
coefficients called Neumann matrices. In this 
section, we will express the three-string vertex,
in the absence of a $B$-background but with zero 
modes included, in terms of a set of new oscillators 
labeled by the eigenvalues of the Neumann matrices. 
The three-string vertex in this form has been 
obtained in ref. \cite{Belov}; our brief review 
here presents a slightly different approach, and 
serves to set up our notations and introduction 
to proceeding to the more complicated case with 
a nonvanishing $B$-background.

The three-string vertex, with zero modes included, 
that defines Witten's star product is given
\ben \label{V0}
|V'_3\rangle =\exp [-{1\over 2}
\sum_{r,s} \sum_{m,n\geq0}
a_m^{M(r)\dag}V'^{rs}_{mn}G_{MN}
a_n^{N(s)\dag} ]|0\rangle\ .
\een
Here the superscripts $r,s=1,2,3$ label different 
strings, the subscripts $m, n =0,1,2,\cdots$ the 
string modes, $G^{MN}=\eta^{MN}=\mbox{diag}(-1,1,\ldots,1)$ 
is the metric of target space, and $a_n^{M(r)\dagger}$ 
the creation operator of the corresponding string mode. 
The explicit form of the Neumann matrices $V'^{rs}$ 
has been given in refs. \cite{GJ,RSZ1,Okuyama}.

The eigen-problem for the Neumann matrices 
$M'^{rs}=CV'^{rs}$ has been solved in ref. 
\cite{FHM1,Belov} where $C_{mn}=(-1)^n \delta_{mn}$ 
is the twist matrix. Let us first review the 
main results of ref. \cite{FHM1} for our 
purposes. It is known that the zero momentum 
sector of the Neumann matrix $M^{rs}$ has a 
continuous spectrum \cite{RSZ}. After including 
the zero modes, however, the spectrum will 
have a continuous branch labeled by a 
continuous parameter $\ka\in (-\infty,+\infty)$ 
and a discrete branch consisting of a single 
point. The eigen-equation reads
\ben \label{eigeneq0}
\sum_{n=0}^{\infty}M'^{rs}_{mn}u_{n}(\ka)
&=&\rho^{rs}(\ka)u_{m}(\ka)\ ,\quad\mbox{for 
the continuous spectrum} 
\label{eec}\\
\sum_{n=0}^{\infty}M'^{rs}_{mn}\phi_{n}
&=&\rho^{rs}_0\phi_{m}\ ,\quad\mbox{for the
discrete spectrum} \label{eed}
\een

The continuous spectrum consists of the open
interval $(-{1\over 3},0)$. \footnote{Though
in the following we are going to quote many 
results from ref. \cite{FHM1}, our terminology
for the continuous and discrete spectra is 
different from theirs. According to the 
definition in ref. \cite{FHM1}, some points 
between $-{1\over3}$ and $0$, at which the 
general expression given in that paper for 
the eigenvectors in the continuous spectrum 
takes the form $0/0$, are said to belong to 
the discrete spectrum. However, a careful 
study of the limit shows that the 
eigenvectors are in fact continuous at 
these points. So we say these points still
belong to the continuous spectrum.}
The expression for the continuous eigenvalues 
of $M'\equiv M'^{11}$ is given by
\be
\label{eigenvc0}
\rho(\ka)\equiv \rho^{11}(\ka) = 
{-1\over 1+2\cosh {\pi\ka \over 2}}\ .
\ee
The eigen-values of $M'^{12}$ and $M'^{21}$ 
can be obtained using properties \refb{A8} in 
Appendix A; they are 
\ben \label{evc12}
\rho^{12} (\ka)
&=&\frac 12 \mbox{sign}
(\ka)\sqrt{[1+3\rho(\ka)][1-\rho(\ka)]}
+\frac 12 [1-\rho(\ka)]\ ,\nonumber \\
\rho^{21}(\ka)
&=&-\frac 12 \mbox{sign}(\ka)\sqrt{[1+3\rho(\ka)]
[1-\rho(\ka)]}+\frac 12 [1-\rho(\ka)]\ ,
\een
and other eigenvalues are given by 
$\rho^{23}=\rho^{31}=\rho^{12}$,
$\rho^{32}=\rho^{13}=\rho^{21}$ 
and $\rho^{22}=\rho^{33}=\rho^{11}$.

The eigenvectors $u_{n}(\ka)$ have following property
\be \label{pp1}
u_{2n}(-\ka)=u_{2n}(\ka),\qquad u_{2n+1}(-\ka)
=-u_{2n+1}(\ka) \ .
\ee
From eq. \refb{eec} we know that, if $u(\ka)$ is an
eigenvector with eigenvalue $\rho^{rs}(\ka)$ then 
$u(-\ka)$ is an eigenvector with eigen-value 
$\rho^{rs}(-\ka)$. It is easy to see that these two 
eigenvectors are degenerate for $M'$ but not for 
$M'^{rs}$ with $r\neq s$. We can restrict to $\ka\geq 0$ 
and construct the following two vectors
\be
(u_+)_n=\left(u_{2n}(\ka)\ ,\ u_{2n+1}(\ka)\right)^t\ ,\quad
(u_-)_n=\left(u_{2n}(\ka)\ ,\ -u_{2n+1}(\ka)\right)^t\ .
\ee
After writing matrices $M'^{rs}$ in the following 
$2\times 2$ block form
\[
\left(
\begin{array}{cc}
M'^{rs}_{ee} & M'^{rs}_{eo}\\
M'^{rs}_{oe} & M'^{rs}_{oo}
\end{array}
\right) 
\]
where $e(o)$ indicate even (odd) modes, the
eigen-equation \refb{eec} now reads
\be
M'^{rs}u_\pm(\ka)
=\rho^{rs}_{\pm}(\ka)u_\pm (\ka) \ ,
\ee
where the summation over $n$ is implied (we 
will use this notation hereafter), and
\be
\rho^{rs}_{\pm}(\ka)=\rho^{rs}(\pm\ka)|_{\ka\geq 0} 
\ee

The discrete spectrum consists of a single point 
in $(0,1)$. The eigenvalue $\rho_0\equiv 
\rho_0^{11}$ is the solution of the following 
equation
\be \label{eigenvd0}
2h(x)=b-4(\gamma+\log(4))\ ,
\ee
where $\gamma$ is the Euler constant, 
$b$ a gauge parameter \cite{RSZ1}, and
\[
h(x)=\psi(-g(x))+\psi(1+g(x))\ ,\quad
g(x)={i\over {2\pi}}\mbox{arcsech}
(-\frac{2x}{1+x})\ ,
\]
with $\psi(x)$ the digamma function. The 
solution to eq. \refb{eigenvd0} in the
interval $(0,1)$ depends on the gauge 
parameter $b$; there always exists only 
one solution no matter what $b$ is. The 
discrete eigen-value for
$M'^{rs}$ with $r\neq s$ is given by
\ben\label{evd012}
\rho^{12}_{0,+}=\rho^{21}_{0,-}
&=&\frac 12 \sqrt{(1+3\rho_0)(1-\rho_0)}
+\frac 12 (1-\rho_0)\ ,\nonumber \\
\rho^{12}_{0,-}=\rho^{21}_{0,+}
&=&-\frac 12 \sqrt{(1+3\rho_0)(1-\rho_0)}
+\frac 12 (1-\rho_0)\ .
\een

Denote the eigenvector of $M'$ at this discrete 
eigenvalue by $\phi_n$. We can construct the
following two vectors the same way as in the 
continuous spectrum
\be
(\phi_+)_n=\left(\phi_{2n}\ ,\ \phi_{2n+1}\right)^t\ ,\quad
(\phi_-)_n=\left(\phi_{2n}\ ,\ -\phi_{2n+1}\right)^t\ ,
\ee
and the eigen-equation \refb{eed} will be rewritten as
\be
M'^{rs}\phi_\pm
=\rho^{rs}_{\pm}\phi_\pm  \ ,
\ee

In this paper, we will use properly 
normalized eigenvectors. So we can write 
down the following completeness and 
ortho-normal relations $(\mbox{with}\ a,b=+,-)$ 
\ben
&& u^{\dag}_a(\ka)u_b(\ka')
=\delta_{ab}\delta(\ka-\ka')\ ,\quad
\phi^{\dag}_a\phi_b
=\delta_{ab} \ , \quad
u^{\dag}_a\phi_b=0 \ ,\\[2mm]
&&\sum_{a=+,-}\left(\int_{0}^{\infty}d\ka\
u_{a}(\ka)u^{\dag}_{a}(\ka)+
\phi_{a}\phi^{\dag}_{a}\right) =Id\ .  
\een
In these relations $\ka, \ka'\geq 0$. 

Now we can rewrite the Neumann matrices in the 
following diagonal form
\be
M'^{rs}=\sum_{a=+,-}\left(\int_{0}^{\infty}d\ka\
\rho^{rs}_a(\ka)_a u_{a}(\ka)u^{\dag}_{a}(\ka)+
\rho_{0,a}^{rs}\phi_{a}\phi^{\dag}_{a}\right)
\ee
Substituting this expression to \refb{V0}, we can 
introduce the following two sets of new oscillators 
to rewrite the three string vertex:
\ben \label{eota0}
&&e^{M\dag}_{\ka}=\sqrt{2}
      \sum_{n=0}^{\infty}u_{2n}(\ka)
a_{2n}^{M\dag}\ ,\quad
o^{M\dag}_{\ka}=-i\sqrt{2}
      \sum_{n=0}^{\infty}u_{2n+1}
(\ka)a_{2n+1}^{M\dag}\ , \nonumber\\
&&\te^{M\dag}=\sqrt{2}
      \sum_{n=0}^{\infty}\phi_{2n}
a_{2n}^{M\dag}\ ,\qquad\;\,
\ot^{M\dag}=-i\sqrt{2}
      \sum_{n=0}^{\infty}\phi_{2n+1}
a_{2n+1}^{M\dag}\ . 
\een
Here, we have used the notation 
$\ \large\tilde{}\ $ to
label quantities associated with the 
discrete eigenvalue, and suppressed the 
string index $r$ in the above equations. 
These new oscillators satisfy the 
commutation relations
\ben
&&[e_\ka^{M}\ ,\  e_{\ka'}^{N\dag} ]
= [o_\ka^{M}\ ,\ o_{\ka'}^{N\dag}]
=G^{MN} \delta(\ka - \ka')\ , \;\;
[e_\ka^{M}\ ,\ o_{\ka'}^{N\dag} ]
= [o_\ka^{M}\ ,\ e_{\ka'}^{N\dag}]
= 0\ ,\nonumber\\[3mm]
&&[\te^M\ ,\ \te^{N\dag}]
= [\ot^M\ ,\ \ot^{N\dag}]
=G^{MN} \ , \qquad\qquad\;\;
[\te^{M}\ ,\ \ot^{N\dag} ]
= [\ot^{M}\ ,\ \te^{N\dag}]
= 0\ ,
\een
and the BPZ conjugation
\ben
bpz(e_\ka^{M}) = -e_\ka^{M\dag}\ ,\quad
bpz(o_\ka^{M}) = -o_\ka^{M\dag}\ ,\nonumber \\[3mm]
bpz(\te^{M}) = -\te^{M\dag}\ ,\quad
bpz(\ot^{M}) = -\ot^{M\dag}\ .
\een

The transformation \refb{eota0} is unitary. 
Its inverse transformation is given by
\ben \label{ieota0}
&&a_{2n}^{M\dag}=\sqrt{2}\left(
\int_{0}^{\infty}d\ka \ u_{2n}(\ka)
e_\ka^{M\dag} +\phi_{2n}
\te^{M\dag} \right)\ , \nonumber\\
&&a_{2n+1}^{M\dag}=i\sqrt{2}\left(
\int_{0}^{\infty}d\ka \ u_{2n+1}(\ka)
o_\ka^{M\dag} +
\phi_{2n+1}\ot^{M\dag}\right)\ .
\een
Finally, we obtain 
the diagonal form of the three-string vertex 
$|V'_3\rangle$:
\ben \label{strvt0}
  |V'_{3}\rangle &=&\exp
\,\Bigl[\, \int_{0}^\infty d\ka\, G_{MN}
\Bigl\{-{1\over 2}\rho(\ka) \, 
\Bigl({e_\ka^{M(1)\dag}} {e_\ka^{N(1)\dag}}
     +{o_\ka^{M(1)\dag}} {o_\ka^{N(1)\dag}} 
+ \hbox{cycl.}
\Bigr)\nonumber\\&&\qquad\qquad\qquad\qquad\quad
- \rho'(\ka)
\Bigl({e_\ka^{M(1)\dag}}  {e_\ka^{N(2)\dag}}
+     {o_\ka^{M(1)\dag}}  {o_\ka^{N(2)\dag}}
+ \hbox{cycl.} \Bigr) \nonumber\\
&&\qquad\qquad\qquad\qquad\quad -i\rho''(\ka)
\Bigl({e_\ka^{M(1)\dag}}  {o_\ka^{N(2)\dag}}
- {o_\ka^{N(1)\dag}}  {e_\ka^{M(2)\dag}}
+ \hbox{cycl.}
\Bigr)\Bigr\}\nonumber \\&&\qquad\qquad\quad 
+ G_{MN} \Bigl\{  -{1\over 2}\rho_0 \,
\Bigl({\te^{M(1)\dag}} {\te^{N(1)\dag}}
     +{\ot^{M(1)\dag}} {\ot^{N(1)\dag}} 
+ \hbox{cycl.} \Bigr)
\\&&\qquad\qquad\qquad\qquad\quad - \rho'_0
\Bigl({\te^{M(1)\dag}}  {\te^{N(2)\dag}}
+     {\ot^{M(1)\dag}}  {\ot^{N(2)\dag}}
+ \hbox{cycl.} \Bigr)
\nonumber\\&&\qquad\qquad\qquad\qquad\quad - i\rho''_0
\Bigl({\te^{M(1)\dag}}  {\ot^{N(2)\dag}}
- {\ot^{N(1)\dag}}  {\te^{M(2)\dag}}
+ \hbox{cycl.} \Bigr) \Bigr\} \Bigr] 
|0\rangle\,.\nonumber
\een
where
\ben
&&\rho'(\ka)=\frac 12 \left(\rho^{12}_+(\ka)
+\rho^{12}_-(\ka)\right)
\ ,\nonumber \quad
\rho''(\ka)=\frac 12 \left(\rho^{12}_+(\ka)
-\rho^{12}_-(\ka)\right) \ ,\nonumber
\\
&&\rho'_0=\frac 12 \left(\rho^{12}_{0,+}+
\rho^{12}_{0,-}(\ka)\right)
\ ,\nonumber \qquad\quad
\rho''_0=\frac 12 \left(\rho^{12}_{0,+}
-\rho^{12}_{0,-}(\ka)\right)
\ .
\een

\section{Three-string vertex in a $B$-background}

Let us turn on a non-vanishing anti-symmetric 
$B$-background in the 1st and 2nd spatial 
dimensions. It is well-known that the effective 
open string metric, $G^{MN}$, and the effective
anti-symmetric noncommutativity parameter 
$\theta^{\alpha\beta}$ between the 1st and 2nd 
coordinates of the string endpoints are, 
respectively, 
\ben \label{met}
G^{MN}&=&\left\{
\begin{array}{rcll}
G^{\mu\nu}&=&
\eta^{\mu\nu} \ , 
& \mbox{for}\quad
\mu,\nu=0,3\ldots,25\ , \\[3mm] 
G^{\alpha\beta}&=&{1\over \xi}
\delta^{\alpha\beta}
={1\over \xi}\mbox{diag}\{1,1\}\ ,
& \mbox{for} \quad \alpha,
\beta=1,2\ , 
\end{array}
\right. \\
\theta^{\alpha\beta}
&=&-\frac{(2\pi\alpha')^2
B}{\xi}\epsilon^{\alpha\beta}\ ,
\nonumber
\een
where $\epsilon^{\alpha\beta}$ 
is the two-by-two anti-symmetric tensor with 
$\epsilon^{12}=1$ and $\xi=1+(2\pi \alpha' B)^2$. 

The three-string vertex will decompose into
\be
|V_3\rangle=|V_{3,\parallel}\rangle 
\otimes|V_{3,\perp}\rangle\ ,
\ee
where $\parallel$ denotes the parallel directions 
$\alpha,\beta= 1, 2$, and 
$\perp$ the transverse directions 
$\mu,\nu = 0,3,\cdots, 25$. The transverse part 
$|V_{3,\perp}\rangle$ will be of the 
same form as given in the last section. 
Here we focus on the parallel directions. 
The oscillator representation for  
$|V_{3,\parallel}\rangle$ can be written as
\ben
&&|V_{3,\parallel}\rangle =\exp \left[-{1\over 2}
\sum_{r,s} \sum_{m,n\geq0}a_m^{\alpha(r)\dag}
{\cal V}^{rs}_{mn,\alpha\beta}
a_n^{\beta(s)\dag}\right]|\td 0\rangle\ .
\een
where the Neumann matrices 
${\cal V}^{\alpha\beta,rs}_{nm}$ 
have been given in \cite{Sugino,KT,BMS}. We 
list the properties of ${\cal V}^{\alpha\beta,rs}_{nm}$
and $\MM^{\alpha\beta,rs}\equiv C{\cal V}^{\alpha\beta,rs}$ in 
Appendix A.

We write the matrices $\MM^{rs}$ in the 
following $4\times 4$ form:
\be
\left(
\begin{array}{cc}
\Gamma^{rs,11}&\ \Gamma^{rs,12} \\ 
\Gamma^{rs,21}&\ \Gamma^{rs,22}
\end{array}
\right)
\ee
where,
\be
\Gamma^{rs,\alpha\beta}=
\left(
\begin{array}{cc}
\MM^{rs,\alpha\beta}_{ee}&\   \MM^{rs,\alpha\beta}_{eo}\\
\MM^{rs,\alpha\beta}_{oe}&\ \MM^{rs,\alpha\beta}_{oo}
\end{array}
\right) 
\ee

In the next subsection, we will review the results 
obtained in ref. \cite{FHM}. Then in subsection 
$3.2$ we will construct the eigenvectors of 
$\MM^{rs}(r\neq s)$, and use the results obtained 
to rewrite the three-string vertex in the final 
subsection.

\subsection{The Spectrum of the Neumann matrix $\MM^{11}$}

The eigenvalue problem for the Neumann matrix $\MM
\equiv \MM^{11}$ has been solved in ref. \cite{FHM,CL}. 
The spectrum also has two branches: a continuous 
branch, labeled by a continuous parameter $\ka \in 
(-\infty,+\infty)$, and a discrete one labeled by a 
discrete parameter $j=1,2$. The eigen-equation is
\be
\sum_{n=0}^{\infty}\sum_{\beta=1}^{2}
\MM^{\alpha\beta}_{mn}v_{n}^{\beta}(\w)
=\lambda(\w)v_{m}^{\alpha}(\w)\ ,
\ee
where $\w=\ka,j$. For our purpose, we write the 
eigenvector as a 4-row vector as 
\be
X(\w)_n=(v^1_{2n}(\w)\ ,\ v^1_{2n+1}(\w)\ ,
\ v^2_{2n}(\w)\ ,\ v^2_{2n+1}(\w))^t \ ,
\ee
and the eigen-equation as
\be \label{Beec}
\MM X(\w)=\lambda (\w)X(\w)\, .
\ee

For the continuous spectrum, the expression for 
the eigenvalues of $\MM$ is only a rescaling of 
that in the absence of the $B$-field, given by
\be
\label{eigenvc}
\lambda(\ka) 
= {1\over \xi}{-1 \over 1 + 2 \cosh {\pi\ka \over 2}}\ .
\ee
and the two degenerate eigenvectors are 
\ben
X(\ka)_n  &=& (v^1_{2n}(\ka)\ ,\ v^1_{2n+1}(\ka)\ ,
             \ v^2_{2n}(\ka)\ ,\ v^2_{2n+1}(\ka))^t \ ,\\[3mm]
Y'(\ka)_n &=& (-v^2_{2n}(\ka)\ ,\ -v^2_{2n+1}(\ka)\ ,
             \ v^1_{2n}(\ka)\ ,\ v^1_{2n+1}(\ka))^t \ .
\een
The components of these two degenerate vectors have 
the following properties: \\
1. Complex conjugation is given by
\be
v_{n}^{1\ast}(\ka)=v_{n}^{1}(\ka)\ ,\quad
v_{n}^{2\ast}(\ka)=-v_{n}^{2}(\ka)\ .
\ee
2. Under $\ka \to -\ka$, we have
\ben \label{evod}
v_{2n}^{1}(-\ka)=v_{2n}^{1}(\ka)\ ,\quad
v_{2n+1}^{1}(-\ka)=-v_{2n+1}^{1}(\ka)\ ,
\nonumber\\[3mm]
v_{2n}^{2}(-\ka)=-v_{2n}^{2}(\ka)\ ,\quad
v_{2n+1}^{2}(-\ka)=v_{2n+1}^{2}(\ka)\ .
\een

These two degenerate eigenvectors are not orthogonal 
to each other. By the Gram-Schmidt procedure, the 
eigenvector which is orthogonal to $X(\ka)$ is given 
by $Y(\ka)=Y'(\ka)-X^\dagger(\ka)Y'(\ka)X(\ka)$. 
After normalization, we write $Y(\ka)$ in the form
of a 4-row vector as
\be
Y(\ka)_n=(\td v_{2n}^1(\ka)\ ,\ \td v_{2n+1}^1(\ka)\ ,
        \ \td v_{2n}^2(\ka)\ ,\ \td v_{2n+1}^2(\ka))^t\, .
\ee
It is easy to check that $\td v^\alpha_n(\ka)$ has 
the following properties: \\
1. Complex conjugation is given by
\be
\td v_{n}^{1\ast}(\ka)=-\td v_{n}^{1}(\ka)\ ,\quad
\td v_{n}^{2\ast}(\ka)=\td v_{n}^{2}(\ka)\ .
\ee
2. Under $\ka \to -\ka$, we have
\ben \label{evod}
\td v_{2n}^{1}(-\ka)=-\td v_{2n}^{1}(\ka)\ ,\quad
\td v_{2n+1}^{1}(-\ka)=\td v_{2n+1}^{1}(\ka)\ ,
\nonumber\\[3mm]
\td v_{2n}^{2}(-\ka)=\td v_{2n}^{2}(\ka)\ ,\quad
\td v_{2n+1}^{2}(-\ka)=-\td v_{2n+1}^{2}(\ka)\ .
\een

Note that $\lambda(\ka)=\lambda(-\ka)$, we have 
the following $4$-fold degenerate eigenvectors 
for $\MM$ in the continuous spectrum,
\ben
(X_e(\ka))_n &=&\frac 12(X(\ka)_n+X(-\ka)_n)
=(v^1_{2n}(\ka)\ ,\ 0\ ,\ 0\ ,\ v^{2}_{2n+1}(\ka))^t \ , \\
(X_o(\ka))_n &=&\frac 12(X(\ka)_n-X(\ka)_n)
=(0\ ,\ v^1_{2n+1}(\ka)\ ,\ v^{2}_{2n}(\ka)\ ,\ 0)^t \ , \\ 
(Y_e(\ka))_n &=&\frac 12(Y(\ka)_n+Y(-\ka)_n)
=(0\ ,\ \td v^1_{2n+1}(\ka)\ ,\ \td v^{2}_{2n}(\ka)\ ,\ 0)^t \ , \\
(Y_o(\ka))_n &=&\frac 12(Y(\ka)_n-Y(-\ka)_n)
=(\td v^1_{2n}(\ka)\ ,\ 0\ ,\ 0\ ,\ \td v^{2}_{2n+1}(\ka))^t \ .
\een

For discrete spectrum, we have two points lying in 
the interval $(0,\frac 1\xi)$, and they are determined 
by the following equation, respectively, with 
$x\equiv\xi\lambda$:
\be \label{eigenvd}
2h(x)=\mp 4B\pi^2\sqrt{\frac{1-x}{1+3x}}
+b-4(\gamma+\log(4))\ ,
\ee
The double degenerate eigenvectors at each point 
$(j=1,2)$ are 
\ben
(X_e(j))_n &=&(v^1_{2n}(j)\ ,\ 0\ ,\ 0\ ,\ v^2_{2n+1}(j))^t\ , \\[3mm]
(X_o(j))_n &=&(0\ ,\ iv^2_{2n+1}(j)\ ,\ -iv^1_{2n}(j)\ ,\ 0)^t\ .
\een
They are obtained by setting $(D_1=i\ ,\ D_2=0)$ and 
$(D_1=0\ ,\ D_2=-1)$ in eq. (6.1) of ref. \cite{FHM}. 
The components satisfy 
$v^{1\ast}_{2n}(j)=v^1_{2n}(j)\ ,\ v^{2\ast}_{2n+1}(j)=-v^2_{2n+1}(j)$.

\subsection{The Spectrum for $\MM^{rs}\ (r\neq s)$}

In this subsection, we will construct the eigenvectors 
for $\MM^{rs}\ (r\neq s)$ explicitly. 

The continuous eigenvalues of $\MM^{rs}\ (r\neq s)$ 
can be obtained in the same way as in Section 2.  
Using the properties \refb{A8} in Appendix A, we have 
\ben
\lambda^{12}(\ka)&=&\frac 12 \mbox{sign}(\ka)
\sqrt{\left({1\over \xi}+3\lambda(\ka)\right)
\left({1\over \xi}-\lambda(\ka)\right)}+   
\frac 12 \left({1\over \xi}-\lambda(\ka)\right)\ , 
\nonumber\\
\lambda^{21}(\ka)&=&-\frac 12 \mbox{sign}(\ka)
\sqrt{\left({1\over \xi}+3\lambda(\ka)\right)
\left({1\over \xi}-\lambda(\ka)\right)}+
\frac 12 \left({1\over \xi}-\lambda(\ka)\right)\ .
\een
Here, sign$(\ka)$ appears because of the requirement 
that the expressions should recover \refb{evc12}
as $B\to 0$.

The corresponding eigenvectors can be constructed 
as follows. We set (for $\ka\geq 0$)
\ben
&&X_+(\ka)=X_e(\ka)+X_o(\ka)\ ,\quad 
X_-(\ka)=X_e(\ka)-X_o(\ka)\ ,\nonumber\\[2mm]
&&Y_+(\ka)=Y_e(\ka)+Y_o(\ka)\ ,\quad 
Y_-(\ka)=Y_e(\ka)-Y_o(\ka)\ .
\een
It is easy to see that $X_+(\ka)\ (Y_+(\ka))$ is 
just the $X(\ka)\ (Y(\ka))$ restricted to $\ka\geq 0$,
and $X_-(\ka)\ (Y_-(\ka))$ the $X(-\ka)\ (Y(-\ka))$ 
restricted to $\ka\geq 0$. For Neumann matrix $\MM^{11}$, 
it is known that $X_+(\ka)\ (Y_+(\ka))$ and 
$X_-(\ka)\ (Y_-(\ka))$ are degenerate since 
$\lambda(-\ka)=\lambda(\ka)$. However, the last 
property is no longer true for eigenvalues of 
the Neumann matrices $\MM^{rs}\ (r\neq s)$;
i.e., $(X_+(\ka), Y_+(\ka))$ and $( X_-(\ka), Y_-(\ka))$ 
are no longer degenerate for $\MM^{rs}\ (r\neq s)$.  

We can rewrite the eigen-equation \refb{Beec} as:
\ben
\MM^{rs}\Psi_\pm(\ka)=\lambda^{rs}_\pm(\ka)
\Psi_\pm (\ka)\ ,
\een
where $\Psi=X,Y$, and
\ben
\lambda^{rs}_{\pm}(\ka)
=\lambda^{rs}(\pm\ka)|_{\ka\geq 0}\ .
\een 

For the discrete spectrum, it can be shown that
the eigenvectors of $\MM^{rs}$ with $r\neq s$ 
are linear combination of $X_e(j)$ and $X_o(j)$; 
i.e.
\ben
\MM^{rs}[X_e(j)+X_o(j)]
=\lambda^{rs}_{j,+}[X_e(j)+X_o(j)]\ ,\\[3mm]
\MM^{rs}[X_e(j)-X_o(j)]
=\lambda^{rs}_{j,-}[X_e(j)-X_o(j)]\ .
\een
The proof is left to Appendix C. From now on we 
will denote $X_{\pm}(j)\equiv X_e(j)\pm X_o(j)$. 
The independent eigenvalues are given by
\ben
&&\lambda^{12}_+(j)=\frac 12
\sqrt{\left({1\over \xi}+3\lambda(j)\right)
\left({1\over \xi}-\lambda(j)\right)}+
\frac 12 \left({1\over \xi}-\lambda(j)\right)\ , \\
&&\lambda^{12}_-(j)=-\frac 12
\sqrt{\left({1\over \xi}+3\lambda(j)\right)
\left({1\over \xi}-\lambda(j)\right)}+
\frac 12 \left({1\over \xi}-\lambda(j)\right)\ , \\[3mm]
&&\lambda^{21}_+(j)=\lambda^{12}_-(j)\ ,
\quad\lambda^{21}_-(j)=\lambda^{12}_+(j)\ .
\een

The ortho-normal and completeness relations are
expressed as: (with $a,b=+,-$)
\ben
&& X_a(\ka)^{\dag} X_b(\ka')= Y_a(\ka)^{\dag} Y_b(\ka')
=\delta_{ab}\delta(\ka-\ka') \ ,\nonumber \\[2mm]
&&X_a(i)^{\dag} X_b(j)=\delta_{ab}\delta_{ij}\ ,\nonumber \\[2mm]
&&X_a(\ka)^{\dag} Y_b(\ka')=X_a(i)^{\dag} X_b(\ka)=
X_a(i)^{\dag} Y_b(\ka)=0\ ,\\[2mm]
&&\sum_{a=+,-}\left\{\int_0^{\infty}d\ka\ \left(
X_a(\ka)X_a^{\dag}(\ka)+Y_a(\ka)Y_a^{\dag}(\ka)\right)+
\sum_{j=1,2}\  
\left(X_a(j)X_a^{\dag}(j)\right)\right\}=Id \ .\nonumber
\een
Again, in these relations $\ka\geq 0$.

\subsection{Diagonal representation of the three-string 
vertex in a B-background}

The three-string vertex is $|V_{3,\pa}\rangle
=\exp\{-\frac 12 V_{\pa}\}|\td 0\rangle$,
where
\be
V_{\pa}=\sum_{r,s} \sum_{m,n\geq 0}a_m^{\alpha(r)\dag}
(C\MM^{rs})_{mn,\alpha\beta}
a_n^{\beta(s)\dag}\, .
\ee
Using the results obtained in the last subsection, 
we can rewrite the Neumann matrices $\MM^{rs}$ in 
the following form
\be
\MM^{rs}=\sum_{a=+,-}\left\{\int_0^\infty d\ka\ 
\lambda_a^{rs}(\ka)
\left[X_a(\ka)X_{a}^{\dag}(\ka)
+Y_a(\ka)Y_{a}^{\dag}(\ka)\right]
+\sum_{j=1,2}\lambda_{j,a}^{rs}
X_a(j)X_{a}^{\dag}(j)\right\}
\ee
Define $(A^{\dag (r)}|=(a^{\dag 1(r)}_0\ ,...\ ,
\ a^{\dag 1(r)}_m\ ,...\ ,\ a^{\dag 2(r)}_0\ ,...\ ,
\ a^{\dag 2(r)}_m\ ,...)=|A^{\dag (r)})^t $, then 
\ben
V_\pa &=&(A^{\dag (r)}|C\MM^{rs}|A^{\dag (s)})\nonumber \\
&=&(A^{\dag (r)}|C\left\{\int_0^\infty d\ka\ \left[
\xi^2\left(\lambda^{rs}_+(\ka)+\lambda^{rs}_-(\ka)\right)
\right.\right.\nonumber \\
&&\left.\left.
\left\{X_e(\ka)X_e^{\dag}(\ka)+X_o(\ka)X_o^{\dag}(\ka)
+Y_e(\ka)Y_e^{\dag}(\ka)+Y_o(\ka)Y_o^{\dag}(\ka)\right\}
\right.\right.\nonumber \\ 
&&\qquad\qquad\qquad\qquad \left.\left. +
\xi^2\left(\lambda^{rs}_+(\ka)-\lambda^{rs}_-(\ka)\right)
\right.\right.\nonumber \\
&&\left.\left.
\left\{X_e(\ka)X_o^{\dag}(\ka)+X_o(\ka)X_e^{\dag}(\ka)
+Y_e(\ka)Y_o^{\dag}(\ka)+Y_o(\ka)Y_e^{\dag}(\ka)\right\}
\right]\right.\nonumber \\ 
&&\left.+
\sum_{j=1,2}\left[
\xi^2\left(\lambda^{rs}_{j,+}+\lambda^{rs}_{j,-}\right)
\left\{X_e(j)X_e^{\dag}(j)+X_o(j)X_o^{\dag}(j)\right\}
\right.\right.\nonumber \\ 
&&\qquad\quad \left.\left. +
\xi^2\left(\lambda^{rs}_{j,+}-\lambda^{rs}_{j,-}\right)
\left\{X_e(j)X_o^{\dag}(j)+X_o(j)X_e^{\dag}(j)\right\}
\right]\right\}|A^{\dag (s)})\ .
\een

Note that
\ben
&&[CX_e(\w)]^t=X_e^{\dag}(\w)\ ,\quad [CX_o(\w)]^t
=-X_o^{\dag}(\w)\ ,\quad
\w=\ka,j\nonumber \\[3mm]
&&[CY_e(\ka)]^t=Y_e^{\dag}(\ka)\ ,\quad [CY_o(\ka)]^t
=-Y_o^{\dag}(\ka)\ .
\een
Let us introduce the following new oscillators:
\ben
&& e_\ka^{\bar1\dag}=\sqrt2 X_e^{\dag}(\ka)|A^{\dag})
=\sqrt2\sum_{n=0}^\infty \left(v^1_{2n}(\ka)a^{1\dag}_{2n}
-v^2_{2n+1}(\ka)a^{2\dag}_{2n+1}\right) \ ,\label{eoB1}\\
&& o_\ka^{\bar1\dag}=-i\sqrt2 X_o^{\dag}(\ka)|A^{\dag})
=-i\sqrt2\sum_{n=0}^\infty\left(v^1_{2n+1}(\ka)
a^{1\dag}_{2n+1}-v^2_{2n}(\ka)a^{2\dag}_{2n}\right) \ ,\\
&& e_\ka^{\bar2\dag}=\sqrt2 Y_e^{\dag}(\ka)|A^{\dag})
=\sqrt2\sum_{n=0}\left(\td v^2_{2n}(\ka)a^{2\dag}_{2n}
-\td v^1_{2n+1}(\ka)a^{1\dag}_{2n+1}\right) \ ,\\
&& o_\ka^{\bar2\dag}=-i\sqrt2 Y_o^{\dag}(\ka)|A^{\dag})
=-i\sqrt2\sum_{n=0}\left(\td v^2_{2n+1}(\ka)
a^{2\dag}_{2n+1}-\td v^1_{2n}(\ka)a^{1\dag}_{2n}\right)\ ,\\
&& \te^{\bar j\dag}=\sqrt2 X_e^{\dag}(j)|A^{\dag})
=\sqrt2\sum_{n=0}^\infty \left(v^1_{2n}(j)a^{1\dag}_{2n}
-v^2_{2n+1}(j)a^{2\dag}_{2n+1}\right) \ ,\\
&& \ot^{\bar j\dag}=-i\sqrt2 X_o^{\dag}(j)|A^{\dag})
=-i\sqrt2\sum_{n=0}^\infty \left(iv^2_{2n+1}(j)
a^{1\dag}_{2n+1}+iv^1_{2n}(j)a^{2\dag}_{2n}\right) \ .
\label{eoB2}
\een

They satisfy the commutation relations
\ben
&&[e_\ka^{\bar\alpha}\ ,\ e_{\ka'}^{\bar\beta\dag} ]
= [o_\ka^{\bar\alpha}\ ,\ o_{\ka'}^{\bar\beta\dag}]
=G^{\bar\alpha\bar\beta} \delta(\ka - \ka')\ , \quad
[e_\ka^{\bar\alpha} \ ,\ o_{\ka'}^{\bar\beta\dag} ]
= [o_\ka^{\bar\alpha} \ ,\ e_{\ka'}^{\bar\beta\dag}]
= 0\nonumber \\[3mm]
&&[\te^{\bar i}\ ,\ \te^{\bar j \dag} ]
=[\ot^{\bar i}\ ,\ \ot^{\bar j\dag}]
=G^{\bar i\bar j}\ ,\qquad\qquad\qquad
[\te^{\bar i}\ ,\ \ot^{\bar j \dag} ]
=[\ot^{\bar i}\ ,\ \te^{\bar j\dag}]=0\ .
\een
and the BPZ conjugation
\ben
&&bpz(e_\ka^{\bar\alpha}) = -e_\ka^{\bar\alpha\dag}\ , \quad
bpz(o_\ka^{\bar\alpha}) = -o_\ka^{\bar\alpha\dag}\ , \nonumber \\[3mm]
&&bpz(\te^{\bar i}) = -\te^{\bar i\dag}\ ,\quad
bpz(\ot^{\bar i}) = -\ot^{\bar i\dag}\ .
\een

The inverse transformation of \refb{eoB1}-\refb{eoB2} is 
\ben \label{ieotaB}
\sqrt 2 |A^{\dag}\Bigr) &=&\int_0^\infty d\ka \ \left[
  X_e(\ka) e_\ka^{1{\dag}}
+iX_o(\ka) o_\ka^{1{\dag}}
+ Y_e(\ka) e_\ka^{2{\dag}}+
i Y_o(\ka) o_\ka^{2{\dag}}\right] \nonumber \\[3mm]
&&+\sum_{j=1,2}\left[
  X_e(j)\te^{j{\dag}}
+iX_o(j)\ot^{j{\dag}}\right] \ .
\een

Finally, the operator $V_{\parallel}$ can be written 
in the following diagonal form:
\ben \label{Bstrvt}
|V_{3,\parallel}\rangle &=&\exp
\,\left[\, \int_{0}^\infty d\ka\, G_{\bar\alpha\bar\beta}  
\Bigl\{ -{1\over 2}\xi\lambda(\ka) \, 
\Bigl({e_\ka^{\bar\alpha(1)\dag}}{e_\ka^{\bar\beta(1)\dag}}
+ {o_\ka^{\bar\alpha(1)\dag}} {o_\ka^{\bar\beta(1)\dag}} 
+ \hbox{cycl.}
\Bigr) \right.\nonumber\\ &&\qquad\qquad\qquad\qquad 
- \xi\lambda'(\ka)
\Bigl({e_\ka^{\bar\alpha(1)\dag}}{e_\ka^{\bar\beta(2)\dagger}}
+ {o_\ka^{\bar\alpha(1)\dag}}{o_\ka^{\bar\beta(2)\dag}}
+ \hbox{cycl.}
\Bigr) \nonumber\\&&\qquad\qquad\qquad\qquad 
- i\xi\lambda''(\ka)
\Bigl({e_\ka^{\bar\alpha(1)\dag}}{o_\ka^{\bar\beta(2)\dag}}
    - {o_\ka^{\bar\beta(1)\dag}}{e_\ka^{\bar\alpha(2)\dag}} 
+ \hbox{cycl.}
\Bigr)\Bigr\}\nonumber \\&&\qquad\qquad+ 
G_{\bar i\bar j}
\Bigl\{-{1\over 2}\xi\lambda_i \, 
\Bigl({\te^{\bar i (1)\dag}} {\te^{\bar j (1)\dag}}
     +{\ot^{\bar i(1)\dag}} {\ot^{\bar j(1)\dag}} 
+ \hbox{cycl.} \Bigr)
\\&&\qquad\qquad\qquad\quad 
- \xi\lambda'_i
\Bigl({\te^{\bar i (1)\dag}}  {\te^{\bar j (2)\dag}}
   + {\ot^{\bar i (1)\dag}}  {\ot^{\bar j (2)\dag}}
+ \hbox{cycl.} \Bigr) 
\nonumber
\\&&\qquad\qquad\qquad\quad\left.
- i\xi\lambda''_i
\Bigl({\te^{\bar i (1)\dag}}{\ot^{\bar j (2)\dag}}
-     {\ot^{\bar j (1)\dag}}{\te^{\bar i (2)\dag}}
+ \hbox{cycl.} \Bigr)
\Bigr\}\right] |\td 0\rangle\ . \nonumber
\een
Here 
\ben
&&\lambda'(\ka)=\frac 12
\left(\lambda^{12}_+(\ka)+\lambda^{12}_-
(\ka)\right)\ ,\nonumber \quad
\lambda''(\ka)=\frac 12
\left(\lambda^{12}_+(\ka)-\lambda^{12}_-
(\ka)\right)\ ,\nonumber \\
&&\lambda'_i=\frac 12
\left(\lambda^{12}_{i,+}+\lambda^{12}_{i,-}
(\ka)\right)\ ,\nonumber \qquad\quad
\lambda''_i=\frac 12
\left(\lambda^{12}_{i,+}-\lambda^{12}_{i,-}
(\ka)\right)\ .  
\een

The explicit form of the three-string vertex in
full 26 dimensions in the presence of a $B$-background 
can be constructed as follows: If in eq. \refb{strvt0} 
we restrict the indices $M,N$ to the transverse 
directions $\mu\nu$, we get the vertex 
$|V_{3,\perp}\rangle$; then $|V_3\rangle 
=|V_{3,\parallel}\rangle \otimes |V_{3,\perp}\rangle$. 
It is easy to see from the eqs. \refb{Bstrvt} and 
\refb{strvt0} that it is of the same form as eq. 
(3.21) in ref. \cite{DLMZ}, which examined only the 
zero-momentum sector with $B=0$, except that now we 
have two sets of oscillators corresponding to the two types of spectra,
continuous and discrete.

\sectiono{Identification of the full Moyal structure} 
\label{s3}

In this section we will present the explicit 
Moyal structure of the three-string vertex, including 
both the Moyal coordinates and corresponding 
noncommutativity parameters. In ref. \cite{DLMZ}, 
the oscillator form of the three-vertex has been 
identified with the Moyal multiplication in an infinite 
commuting set of two-dimensional noncommutative subspaces. Generalizing their result, in our present case we will 
obtain the Moyal structure for the three-string vertex, 
which is of the following form:
\ben
[x^M(\w),y^N(\w')]_{\star}&=& i \theta_M(\w) G^{MN}
\delta(\w-\w'), \, \nonumber \\
&& (M,N= 0,\ldots,25\ ;\;\;
\w,\w' = \ka \;{\rm or}\; j)
\een
where the space-time metric is the modified one 
given in eq. \refb{met}. 

According to ref. \cite{DLMZ}, if to each pair of 
Moyal coordinates (with $M$ fixed), we associate 
one pair of oscillators:
\be
(x^{M(r)}, y^{M(r)}) \,
\leftrightarrow \,( a^{M(r)\dag}, 
b^{M(r)\dag}) \ , \ r=1,2,3
\ee
then the three-string vertex will be of the 
following form: (no summation over $M$) 
\ben
\label{Moy1}
|V_3(\theta)\rangle &\sim&
\exp \Bigl[ - {1\over 2}
 \Bigl( {-4 + \theta^2\over 12 + \theta^2}\Bigr)
 G^{MM}(a_{M}^{(1)\dag} a_{M}^{(1)\dag}
+ b_{M}^{(1)\dag} b_{M}^{(1)\dag} 
+ \hbox{cycl.} ) \nonumber \\
&&\qquad -\Bigl( 
{8\over 12+ \theta^2} \Bigr) \,\,
G^{MM}(a_{M}^{(1)\dag} a_{M}^{(2)\dag} 
+ b_{M}^{(1)\dag} b_{M}^{(2)\dag} 
+ \hbox{cycl.} )\\
&&\qquad -\Bigl( 
{4i\theta\over 12+ \theta^2} \Bigr)\,\,
G^{MM}(a_{M}^{(1)\dag} b_{M}^{(2)\dag} -  
b_{M}^{(1)\dag} a_{M}^{(2)\dag}
+ \hbox{cycl.} )\Bigr] |0\rangle \ .\nonumber
\een
Here we have suppressed the index $M$ in the 
noncommutativity parameter $\theta_M$ between 
$(x^{M}\ ,\ y^{M})$. Comparing the exponents in 
eqs. \refb{strvt0}, \refb{Bstrvt} with those 
in eq. \refb{Moy1}, we see that 
one should identify
\ben
&&(a^{\mu\dag}\ ,\ b^{\mu\dag}) \leftrightarrow 
(e_\ka^{\mu\dag}\ ,\ o_\ka^{\mu\dag})\, 
\;\;\;\mbox{or}\;\;\; (a^{\mu\dag}\ ,\ b^{\mu\dag}) 
\leftrightarrow (\te^{\mu\dag}\ ,\ \ot^{\mu\dag})\ ,  \\[3mm]
&&(a^{\bar\alpha\dag}\ ,\ b^{\bar\alpha\dag}) 
\leftrightarrow
(e_\ka^{\bar\alpha\dag}\ ,\ o_\ka^{\bar\alpha\dag})\,
\;\;\;\mbox{or}\;\;\; (a^{\bar i\dag}\ ,\ b^{\bar i\dag})
\leftrightarrow (\te^{\bar i\dag}\ ,\ \ot^{\bar i\dag})\, ,
\een
by requiring the following conditions:\\
For the transverse directions $M=\mu$,
\ben 
\begin{array}{ll}
\rho(\ka) =\displaystyle
{-4 + \theta^2_\pe(\ka)\over 12 + \theta^2_\pe(\ka)} \,,
&\rho_0 =\displaystyle  
{-4 + \theta^2_\pe\over 12 + \theta^2_\pe } \ , \\[5mm]
\rho'(\ka)=\displaystyle
{8\over 12+ \theta^2_\pe(\ka)}
\,,&
\rho'_0=\displaystyle{8\over 12+ \theta^2_\pe}
\ ,\\[5mm]
\rho''(\ka)  =\displaystyle
{4\theta_\pe(\ka)\over 12+\theta^2_\pe(\ka)} \,,
&\rho''_0  =\displaystyle
{4\theta_\pe\over 12+\theta^2_\pe}\ .
\end{array}
\een
For the parallel directions $M=\alpha$, 
\ben
\begin{array}{ll}
\xi\lambda(\ka) =\displaystyle
{-4 + \theta^2_\pa(\ka)\over 12 
+\theta^2_\pa(\ka)} \,,
&\xi\lambda_i =\displaystyle
{-4 + \theta^2_{\pa,i}\over 12 
+ \theta^2_{\pa,i}} 
\ ,\\[5mm]
\xi\lambda'(\ka)=\displaystyle
{8\over 12+ \theta^2_\pa(\ka)}
\,,& 
\xi\lambda'_i=\displaystyle
{8\over 12+ \theta^2_{\pa,i}} \ ,\\[5mm]
\xi\lambda''(\ka)  =\displaystyle
{4\theta_\pa(\ka)\over 12
+\theta^2_\pa(\ka)} \,,
&\xi\lambda''_i  =\displaystyle
{4\theta_{\pa,i}\over 12+\theta^2_{\pa,i}}\ . 
\end{array}
\een
This identification also requires the 
consistency conditions
\be
\xi\lambda + \xi\lambda^{12} 
+ \xi\lambda^{21} =1=
\rho + \rho^{12} + \rho^{21}\ .
\ee  
Indeed, these conditions are satisfied because of
the properties \refb{A5} in Appendix A. 

We thus identify the two types of noncommutativity 
parameters as follows: 
\ben\label{thetamoyal}
&&\theta_\pa(\ka) =\theta_\pe(\ka)
= 2\,\tanh \Bigl({\pi\ka\over4}\Bigr)
\ ,\\
&&\theta_\pe=
2\sqrt{\frac{1+3\rho_0}{1-\rho_0}}\ , \quad
\theta_{\pa,1}=
2\sqrt{\frac{1+3\xi\lambda_1}{1-\xi\lambda_1}}\ ,\;\;
\theta_{\pa,2}=-
2\sqrt{\frac{1+3\xi\lambda_2}{1-\xi\lambda_2}}\ ,
\een
where $\rho_0$ is determined by eq. \refb{eigenvd0} 
and $\lambda_j$ is determined by eq. \refb{eigenvd}. 
We see that the two types of noncommutativity 
parameters belong to non-overlapping regions: 
$\theta_{\pe}(\ka)\ , \ \theta_{\pa}(\ka) \in (0,2]$, 
while $\theta_{\pe}\ , \ \theta_{\pa,1} \in 
(2,\infty)\ ,\ \theta_{\pa,2} \in (-\infty,-2)$. Note 
that only $\theta_{\pa,j}$ depends on the $B$-background.  
 
Having obtained the noncommutativity parameters, 
we now consider the Moyal coordinates. Define the 
following coordinate and momentum operators in 
terms of our new oscillators by
\ben
\hat x=\frac {i}{\sqrt 2}(e-e^{\dag} )\ , \qquad 
\hat q_x=\frac {1}{\sqrt 2}(e+e^{\dag} )\ ,\nonumber \\
\hat y=\frac {i}{\sqrt 2}(o-o^{\dag} )\ , \qquad
\hat q_y=\frac {1}{\sqrt 2}(o+o^{\dag} )\ .
\een 

Recall that the string mode expansion is : 
(choosing $\alpha'= {1\over 2}$)
\ben \label{modeexp} 
\widehat X^{\mu}(\sigma)&=&
\  \hat x_0^{\alpha} 
+\sqrt{2} \sum_{n=1}^\infty \hat x_n^{\alpha} \ 
\cos n\sigma\, \nonumber \\
\widehat X^{\alpha} (\sigma)
&=&\  \hat x_0^{\alpha} +\frac{1}{\pi}
\theta^{\alpha\beta}
(\sigma-\frac{\pi}{2})\hat p_{0,\beta}+
\sqrt{2} \sum_{n=1}^\infty [\hat x_n^{\alpha} \,
\cos n\sigma\,+\frac{1}{\pi}\theta^{\alpha\beta}\hat
p_{n,\beta}\,\sin n\sigma]\,,\nonumber \\
\pi\widehat P^{M}(\sigma)&=&\ \hat p_0^{M}+\sqrt{2}
\sum_{n=1}^\infty \hat p_n^{M} \, \cos n\sigma\, ,
\een
where
\ben \label{xoscposc} 
&&\hat x_n^M = {i\over \sqrt{2n}} \,
(a_n^M - a_n^{M\dag}) \,,\qquad 
\hat p_n^M =  \sqrt{n\over2}  \, 
(a_n^M + a_n^{M\dag}) \,,\qquad  n > 0\,, \nonumber \\
&&\hat x_0^M = i\frac{\sqrt{b}}{2} \,  
(a_o^M - a_0^{M\dag}) \,,\qquad
\hat p_0^M =  \sqrt{1\over{b}}  \,
(a_0^M + a_0^{M\dag}) \ .
\een

Making use of the transformations \refb{ieota0}
and \refb{ieotaB}, we can express the coordinate 
operators 
$\{\hat x^\mu_\ka\ ,\ \hat y^\mu_\ka\ ;
\ \hat{\td x}^\mu,\ \hat{\td y}^\mu\}$ and
$\{\hat x^{\bar \alpha}_\ka\ ,\ \hat y^{\bar \alpha}_\ka\ ;
\ \hat{\td x}^{\bar \alpha},\ \hat{\td y}^{\bar \alpha}\}$ 
in terms of the original $\hat x^M_n, 
\hat p^M_n$:
\ben \label{Moyalco}
&&\hat {x}^\mu_\ka=\sqrt{2}
\sum_{n=0}^{\infty}u_{2n}(\ka)
\sqrt{2\bar n}\hat x_{2n}^\mu\ ,\quad
\hat y_{\ka}^\mu=-\sqrt{2}\sum_{n=0}^{\infty}
\frac{u_{2n+1}(\ka)}{\sqrt{2n+1}} 
\hat p_{2n+1}^\mu\ ,\label{416}\\
&&\hat {\td x}^\mu=\sqrt{2}
\sum_{n=0}^{\infty}\phi_{2n}
\sqrt{2\bar n}\hat x_{2n}^\mu\ ,\quad
\hat {\td y}^\mu=-\sqrt{2}\sum_{n=0}^{\infty}
\frac{\phi_{2n+1}}{\sqrt{2n+1}}   
\hat p_{2n+1}^\mu\ ,\\ 
&&\hat {x}^{\bar 1}_{\ka}=\sqrt{2}\sum_{n=0}^{\infty}
\left(v_{2n}^1(\ka)\sqrt{2\bar n}\hat x_{2n}^1
+i\frac{v^2_{2n+1}(\ka)}{\sqrt{2n+1}}\hat p^2_{2n+1}\right) \
,\label{417}\\
&&\hat {y}^{\bar 1}_{\ka}=\sqrt{2}\sum_{n=0}^{\infty}
\left(-\frac{v_{2n+1}^1(\ka)}{\sqrt{2n+1}}\hat p_{2n+1}^1
+iv^2_{2n}(\ka)\sqrt{2\bar n}\hat x^2_{2n}\right) \ ,\\
&&\hat {x}^{\bar 2}_{\ka}=\sqrt{2}\sum_{n=0}^{\infty}
\left(\td v_{2n}^2(\ka)\sqrt{2\bar n}\hat x_{2n}^2
+i\frac{\td v^1_{2n+1}(\ka)}{\sqrt{2n+1}}\hat p^1_{2n+1}\right) \ ,\\
&&\hat {y}^{\bar 2}_{\ka}=\sqrt{2}\sum_{n=0}^{\infty}
\left(-\frac{\td v_{2n+1}^2(\ka)}{\sqrt{2n+1}}\hat p_{2n+1}^2
+i\td v^1_{2n}(\ka)\sqrt{2\bar n}\hat x^1_{2n}\right) \ ,\\
&&\hat {\td x}^{\bar j}=\sqrt{2}\sum_{n=0}^{\infty}
\left(v_{2n}^1(j)\sqrt{2\bar n}\hat x_{2n}^1
+i\frac{v^2_{2n+1}(j)}{\sqrt{2n+1}}\hat p^2_{2n+1}\right) \ ,\\
&&\hat {\td y}^{\bar j}=\sqrt{2}\sum_{n=0}^{\infty}
\left(-i\frac{v_{2n+1}^2(j)}{\sqrt{2n+1}}\hat p_{2n+1}^1
+v^1_{2n}(j)\sqrt{2\bar n}\hat x^2_{2n}\right) \ .\label{422}
\een
where $\bar n=\frac{1}{b},\ \mbox{for}\ n=0$. We 
immediately see that in the directions parallel to 
the background $B$-field, the string modes in different 
spatial directions get mixed with each other. 

The Moyal coordinates $\{{x}^\mu(\ka)\ ,\ {y}^\mu(\ka)\ ; 
\ {\td x}^\mu\ ,\ {\td y}^\mu\}$ and
$\{{x}^{\bar\alpha}(\ka)\ ,\ {y}^{\bar\alpha}(\ka)\ ; 
\ {\td x}^{\bar\alpha}\ ,\ {\td y}^{\bar\alpha}\}$
are the eigenvalues of operators 
$\{\hat{x}_{\ka}^\mu\ ,\ \hat{y}_{\ka}^\mu\ ; 
\ \hat{\td x}^\mu\ ,\ \hat{\td y}^\mu \}$ and
$\{\hat{x}_{\ka}^{\bar\alpha}\ ,\ \hat{y}_{\ka}^{\bar\alpha}\ ; 
\hat{\td x}^{\bar\alpha}\ ,\ \hat{\td y}^{\bar\alpha}\}$ 
respectively. Their eigenvectors can be constructed by
starting from the eigenvectors of $\hat x^M_n, 
\hat p^M_n$:
\ben
&&\prod_{n}|x_{n}^{M}\rangle
=\exp\left\{-{1\over 2}(x|H^2|x) 
-\sqrt{2}i(a^{\dag}|H|x)
+{1\over 2}(a^{\dag}|a^{\dag})
\right\}|0\rangle\ ,\\
&&\prod_{n}|p_{n}^{M}\rangle
=\exp\left\{-{1\over 2}(p|H^{-2}|p) 
+\sqrt{2} (a^{\dag}|H^{-1}|p)
-{1\over2}(a^{\dag}|a^{\dag})
\right\}|0\rangle\ .
\een
where $H_{mn}=\delta_{mn}\sqrt{n}+ 
\delta_{m0}\delta_{n0}\sqrt{2\over b}$
and $(z|f|z)=\sum\limits_{m,n\geq0}z_m^M f_{mn} 
z_{n,M}$ (no summation over $M$). 
From eqs. \refb{416}-\refb{422}, we see that only  
$x_{2n}$ and $p_{2n+1}$ are 
involved in the Moyal coordinate operators. Define 
\ben
&&{X}^{\mu}(\ka)=({x}^\mu(\ka)\ ,\ {y}^\mu(\ka))\ ,\quad
{E}_{\ka}^{\mu}=(e_\ka^{\mu\dag}\ ,\ o_\ka^{\mu\dag})\ ,\\
&&{\td X}^{\mu}=({\td x}^\mu\ ,\ {\td y}^\mu)\ ,\qquad\quad
\qquad {\td E}^{\mu}=(\te^{\mu\dag}\ ,\ \ot^{\mu\dag})\ ,\\
&&{X}^{\bar\alpha}(\ka)=({x}^{\bar\alpha}(\ka)\ ,
\ {y}^{\bar\alpha}(\ka))\ ,\quad
{E}_{\ka}^{\bar\alpha}=(e_\ka^{\bar\alpha\dag}\ ,
\ o_\ka^{\bar\alpha\dag})\ ,\\
&&{\td X}^{\bar\alpha}=({\td x}^{\bar\alpha}\ ,
\ {\td y}^{\bar\alpha})\ ,\qquad\qquad\quad
{\td E}^{\bar\alpha}=(\te^{\bar\alpha\dag}\ ,
\ \ot^{\bar\alpha\dag})\ ,
\een
then the eigenstates of the Moyal coordinates are 
\ben
|x^\mu(\ka), y^\mu(\ka)\rangle
&=&\exp\left\{\int_0^\infty d\ka\left(
-{1\over 2}{X}(\ka)^{\mu}\cdot
{X}^t_{\mu}(\ka)\right.\right.\nonumber \\
& &\qquad\left.\left. +\sqrt{2}
i{E}_{\ka}^{\mu}\cdot {X}^t_{\mu}(\ka)
 +{1\over 2}{E}_{\ka}^{\mu}\cdot 
{E}^t_{\ka \mu}\right)\right\} |0\rangle\ ,\\
|{\td x}^\mu, {\td y}^\mu\rangle
&=&\exp\left\{
\left( -{1\over 2}{\td X}^{\mu}\cdot 
{\td X}^t_{\mu}\right.\right.\nonumber\\
 &&\qquad
\left.\left.+\sqrt{2}i{\td E}^{\mu}\cdot 
{\td X}^t_{\mu} +{1\over 2}{\td E}^{\mu}\cdot 
{\td E}^t_{\mu}\right)\right\}
|0\rangle\ ,\\
|x^{\bar\alpha}(\ka), y^{\bar\alpha}(\ka)\rangle
&=&\exp\left\{\int_0^\infty d\ka\left(
-{1\over 2}{ X}(\ka)^{\bar\alpha}\cdot
{X}^t_{\bar\alpha}(\ka)\right.\right.\nonumber \\
& &\qquad\left.\left. +\sqrt{2}
i{E}_{\ka}^{\bar\alpha}\cdot {X}^t_{\bar\alpha}(\ka)
 +{1\over 2}{E}_{\ka}^{\bar\alpha}\cdot 
{E}^t_{\ka \bar\alpha}\right)\right\} |0\rangle\ ,\\
|\td x^{\bar\alpha}, \td y^{\bar\alpha}\rangle
&=&\exp\left\{
\left( -{1\over 2}{\td X}^{\bar\alpha}\cdot 
{\td X}^t_{\bar\alpha}\right.\right.\nonumber\\
 &&\qquad
\left.\left.+\sqrt{2}i{\td E}^{\bar\alpha}\cdot 
{\td X}^t_{\bar\alpha} +{1\over 2}{\td E}^{\bar\alpha}
\cdot {\td E}^t_{\bar\alpha}\right)\right\}
|0\rangle\ .
\een

Finally we are in a position to write down the 
commutation relations for the Moyal coordinates: 
(for $\mu,\nu = 0,3,\cdots,25;\ \alpha,\beta = 1,2;\ $ 
and $\ka>0$)
\ben
\label{motheta}
&&[{x}^{\mu}(\ka)\ , \ {y}^{\nu}(\ka')]_\ast
=i2\tanh{\frac{\pi\ka}{4}}G^{\mu\nu}\delta(\ka-\ka')\ ,
\label{motheta1}\\
&&[{\td x}^{\mu}\ ,\  {\td y}^{\nu}]_\ast
=i2\sqrt\frac{1+3\rho_0}{1-\rho_0}
G^{\mu\nu}\ , 
\label{motheta2}\\
&&[{x}^{\bar\alpha}(\ka)\ ,
\ {y}^{\bar\beta}(\ka')]_\ast
=i2\tanh{\frac{\pi\ka}{4}}G^{\bar\alpha\bar\beta}
\delta(\ka-\ka')\ ,\label{motheta3}\\
&&[{\td x}^{\bar 1}\ ,\  
{\td y}^{\bar 1}]_\ast 
=i2\sqrt\frac{1+3\xi\lambda_1}
{1-\xi\lambda_1}G^{\bar 1\bar 1}\ , 
\label{motheta4} \\
&&[{\td x}^{\bar 2}\ ,\
{\td y}^{\bar 2}]_\ast 
=-i2\sqrt\frac{1+3\xi\lambda_2}
{1-\xi\lambda_2}G^{\bar 2\bar 2}\ . 
\label{motheta5}
\een

Let us discuss our results at $\ka=0$. We have 
already seen from ref. \cite{DLMZ} that in the 
zero momentum sector without $B$-background only 
the twist even eigenvectors survive, this makes 
two noncommutative coordinates collapse to a 
commutative one, which was interpreted as the 
momentum carried by half of the string. In our
present case, we have the following remarks: 
 
1. For Neumann matrices with zero modes, 
according to ref. \cite{FHM1}, when $\kappa=0$, 
the determinant defined at this point is zero, 
so the eigenvector should be modified. They 
found two eigenvectors which are denoted as 
$u_{\pm,-{1\over3}}$. However, it was pointed 
out in ref. \cite{Belov} that the double 
degeneracy at $-1/3$ in ref. \cite{FHM1} is due 
to an improper normalization. At $-1/3$, the 
degeneracy of eigenvector should be one which 
is $u_{+,-{1\over3}}$. The explicit form of 
$u_{+,-{1\over3}}$ shows that in the perpendicular
dimensions, only $\hat{x}^{\mu}$ does not vanish;
so only one Moyal coordinate for each perpendicular
dimension will survive. The physical meaning of 
this coordinate is that it is proportional to the 
midpoint coordinate of string $X^\mu (\frac \pi 2)$ 
\cite{DLMZ,Belov}.
      
2. With a $B$-background turned on, we have
\be
v_{2n}^1(0)=0,\qquad v_{2n+1}^2(0)=0,
\ee
at $\kappa=0$. Thus, in the parallel dimensions,
only $\hat{y}^{\bar\alpha}$ survives, resulting 
in two commuting coordinate because of two spatial 
dimensions.

\section{The large-$B$ contraction}

Witten \cite{EW2} has pointed out that the study of 
noncommutative tachyon condensation is considerably 
simplified in the large-$B$ limit. This is because 
in this limit the string field star algebra 
factorizes into two commuting sub-algebras as
${\cal A}={\cal A}_0 \otimes {\cal A}_1$. Here 
${\cal A}_1$ is the algebra of functions in the
noncommutative directions that acts only on the 
string center of mass, while ${\cal A}_0$ is the 
string field algebra in the zero-momentum sector 
if the $B$-filed has the maximal rank. If $B$ has 
less than maximal rank, then ${\cal A}_0$ may carry
momentum, but only in commutative directions. 
Previously this statement has been proved in terms 
of either vertex operators \cite{EW2} or oscillator 
modes \cite{Schnabl}. Here we demonstrate this 
large-$B$ contraction explicitly in our Moyal
representation \refb{motheta1}-\refb{motheta5}
of the string field algebra. 

Suppose the 1st and 2nd spatial dimensions are the 
only noncommutative directions. The key issue for 
the large-$B$ contraction is the fate of the {\it 
two} discrete Moyal pairs in the two directions. 
(The commutation relations of continuous Moyal 
pairs are always the same as in the zero-momentum 
sector.) Certainly only one discrete Moyal pair 
gives rise to the center-of-mass noncommutative
function algebra. We need to identify this pair
and examine what happens to the other discrete 
Moyal pair? We will show that the latter simply 
drops out the contraction, because of the singular 
behavior of the eigenvectors of the Neumann matrices 
at the corresponding eigenvalue: They simply vanish 
in the large-$B$ limit!   

Suppose the $B$-field does not vanish only in the 
1st and 2nd spatial dimensions; we rewrite the 
string mode expansion in these two directions 
(eq. \refb{modeexp}) as
\be
\widehat X^{\alpha} (\sigma)
=\ \hat x^{\alpha} +\frac{\sigma}{\pi}
\theta^{\alpha\beta}
\hat p_{\beta}+
\sqrt{2} \sum_{n=1}^\infty [\hat x_n^{\alpha} \,
\cos n\sigma\,+\frac{1}{\pi}\theta^{\alpha\beta}\hat
p_{n,\beta}\,\sin n\sigma]\,,
\ee
where
\ben
\hat x^{\alpha}&=&\hat x^{\alpha}_0-\frac12
\theta^{\alpha\beta}\hat p_{0,\beta}\ ,\nonumber \\
\hat p^{\alpha}&=&\hat p^{\alpha}_0
\een
The commutation relations for the Fock space 
generators are
\ben \label{Fockcc}
[x_m^{\alpha}\ ,\ p_n^{\beta}]&=& iG^{\alpha\beta}\delta_{mn}\ , \quad 
m,n\ge 1 \nonumber \\
{[\hat x^{\alpha}\ ,\ \hat p^{\beta} ]} 
&=& iG^{\alpha\beta}\ , 
\nonumber \\
{[\hat x^{\alpha}\ ,\ \hat x^{\beta} ]} 
&=& i\theta^{\alpha\beta} \ .
\een

Set $B=t\ B_0$ and take the limit $t\to \infty$. 
The open string parameters $G^{\alpha\beta}\ ,\theta^{\alpha\beta}$ 
scale like
\be
G^{\alpha\beta}\sim G^{\alpha\beta}_0\ t^{-2}\ ,
\qquad \theta^{\alpha\beta}\sim 
\theta^{\alpha\beta}_0 \ t^{-1}\ .
\ee 
In order to get the contraction of the string field 
algebra, we should rescale the Fock space generators 
to make the commutation relations \refb{Fockcc} 
have a definite limit as $B\to \infty$; so one
chooses \cite{Schnabl}
\ben
&&\hat x^{\alpha}_m \to \hat x^{\alpha}_m \ t^{-1}\ ,
\quad \hat p^{\alpha}_m \to \hat p^{\alpha}_m \ 
t^{-1}\ , \qquad m\geq 1 
\nonumber \\[3mm]
&&\hat x^{\alpha} \to \hat x^{\alpha} \ t^{-\frac1 2}
\ ,\quad \hat p^{\alpha} \to \hat p^{\alpha} \ 
t^{-\frac3 2} .
\een

Note that our Moyal coordinates are transformed from 
string modes using eigenvectors of Neumann matrices, 
we should also investigate the behavior of these 
eigenvectors as $B\to \infty$. For the eigenvectors 
in the continuous spectrum,
\ben
&&v_0^1(\ka)\sim t^{-4}\ ,\qquad v_n^1(\ka)\sim 
|\ka\rangle_n\ , \nonumber \\
&&v_0^2(\ka)\sim t^{-3}\ ,\qquad v_n^2(\ka)
\sim t^{-3}\frac{1}{\rho(\ka)-M}
(|v_{e}\rangle+|v_{o}\rangle)_n\ ,
\nonumber \\
&& \td v_0^1(\ka)\sim t^{-3}\ ,\qquad 
\td v_n^1(\ka)\sim t^{-3}\frac{1}{\rho(\ka)-M}
(|v_{e}\rangle+|v_{o}\rangle)_n\ ,
\nonumber \\
&& \td v_0^2(\ka)\sim t^{-4}\ ,\qquad 
v_n^2(\ka)\sim |\ka\rangle_n\ ,
\een
where $|\ka\rangle$ is the continuous eigenvector 
of the matrix $M$ \cite{RSZ}
and the notations $|v_{e,o}\rangle$ are 
given in Appendix A. For the eigenvectors in the discrete 
spectrum, there are some subtleties. It is easy to 
show that as $B\to \infty$, one eigenvalue approaches 
to unity, while the other to zero as $e^{-e^B}$. 
It can be shown that the eigenvector with eigenvalue 
approaching to zero will become vanishing as 
$B\to \infty$. We leave the proof to Appendix B.2. 
The eigenvector with eigenvalue approaching to unity 
scales like
\ben
&&v_0^1(1)\sim 1\ ,\qquad v_n^1(1)
\sim t^{-2}\frac{1}{1-M}|v_{e}\rangle_n\ , \nonumber \\
&&v_0^2(1)=0\ ,\qquad v_n^2(1)
\sim t^{-1}\frac{1}{1-M}|v_{o}\rangle_n\ ,
\een
Thus only one Moyal pair
of coordinates in the discrete spectrum survives
the large-$B$ contraction. 

Keeping only the leading order in $t$, the eqs. 
\refb{417}-\refb{422} change to
\ben
&&\hat x_{\ka}^{1}=t^{-1}\sqrt{2}\sum_{n=1}^{\infty}|\ka\rangle_{2n}
                      \sqrt{2\bar n}\hat x_{2n}^1 \ ,\qquad
\hat y_{\ka}^{1}=-t^{-1}\sqrt{2}\sum_{n=1}^{\infty}
           \frac{|\ka\rangle_{2n+1}}{\sqrt{2n+1}}\hat p_{2n+1}^{1}\ ,\\
&&\hat x_{\ka}^{2}=t^{-1}\sqrt{2}\sum_{n=1}^{\infty}|\ka\rangle_{2n}
                      \sqrt{2\bar n}\hat x_{2n}^2 \ ,\qquad
\hat y_{\ka}^{2}=-t^{-1}\sqrt{2}\sum_{n=1}^{\infty}
           \frac{|\ka\rangle_{2n+1}}{\sqrt{2n+1}}\hat p_{2n+1}^{2}\ ,\\
&&\hat {\td x}^{\bar 1}=t^{-\frac1 2}\frac{2}{\sqrt b}\hat x^1_0\ ,\qquad
\hat {\td y}^{\bar 1}=t^{-\frac1 2}\frac{2}{\sqrt b}\hat x^2_0\ ,
\een

Now, we are ready to read off the large $B$ contraction 
of our Moyal representation of string field algebra 
\refb{motheta1}-\refb{motheta5} 
(for $\mu,\nu = 0,3,\cdots,25;\ \alpha,\beta = 1,2;\ $
and $\ka>0$):
\ben
&&[{x}^{\mu}(\ka)\ , \ {y}^{\nu}(\ka')]_\ast
=i2\tanh{\frac{\pi\ka}{4}}G^{\mu\nu}\delta(\ka-\ka')
\ , \label{fact1}\\
&&[{\td x}^{\mu}\ ,\  {\td y}^{\nu}]_\ast
=i2\sqrt\frac{1+3\rho_0}{1-\rho_0}G^{\mu\nu}\ ,
\label{fact2}\\
&&[{x}^{\alpha}(\ka)\ ,\ {y}^{\beta}(\ka')]_\ast
=i2\tanh{\frac{\pi\ka}{4}}G_0^{\alpha\beta}
\delta(\ka-\ka')\ , \label{fact3}\\
&&[x_0^1\ ,\ x_0^2]_\ast=i\frac12\ .\label{fact4}
\een
The commutation relations \refb{fact1}-\refb{fact3} 
are those for the sub-algebra ${\cal A}_0$ and 
those \refb{fact4} the sub-algebra ${\cal A}_1$.
We note that after recovering string tension, the
noncommutativity parameter for the center-of-mass
function algebra is nothing but $\alpha'$.

\sectiono{Summary of results and discussions}

In this paper, we have found that the 
commuting Moyal-pair structure of Witten's 
star algebra for open string fields persists 
in the presence of a constant $B$-background. 
We have worked out explicitly a new basis in 
which Witten's three-string vertex for string 
fields is diagonalized, and identified the 
commuting Moyal pairs and the corresponding 
noncommutativity parameters. The full set of 
commutation relations for Witten's star algebra 
in the Moyal representations are summarized by 
eqs. \refb{motheta1}-\refb{motheta5}.  

A central issue is the interplay between the 
noncommutativity due to three-string overlap 
and that due to a background $B$-field. It is 
known \cite{DLMZ} that in the zero-momentum 
sector, i.e. without string zero-modes, the 
commuting Moyal pairs for Witten's star algebra 
are labeled by a continuous parameter $\ka 
\in [0,\infty]$. If the strong zero-modes are 
included, then besides those commuting ones
labeled by $\ka$ there is an extra Moyal pair 
commuting with them \cite{Belov}. In our present 
case, in the presence of a $B$-background, the
situation in the commutative directions is the 
same as before. But in each pair of noncommutative 
directions in which the block-diagonal $B$-field 
has non-vanishing components, besides the commuting 
Moyal pairs  labeled by $\ka$ there are {\it two} 
extra Moyal pairs commuting with them and with each 
other. As for noncommutativity parameters, only those
between the discrete Moyal pairs are $B$-dependent, 
while the Moyal parameters between the continuous
pairs are the same as the case in the zero momentum
sector. However, we note that the transformations, 
eqs. \refb{417}-\refb{422}, from the oscillator 
modes to the Moyal pairs are $B$-dependent both 
for the continuous and discrete ones. 

Moreover, we have studied the large-$B$ contraction
for Witten's star algebra. Indeed, we have confirmed 
Witten's statement \cite{EW2}, in the Moyal 
representation, that the large-$B$ contraction 
consists of two commuting sub-algebras: one is the 
ordinary noncommutative function algebra for the 
center of mass of the string in the noncommutative 
directions, while the other commuting sub-algebra
is the same as the star algebra in the zero momentum
sector in the noncommutative direction, and allows
to carry momentum in commutative directions. Our 
contribution to this topic is that we have clarified
the fate of the other discrete Moyal pair in the 
noncommutative directions: It drops out the 
contraction, because the singular behavior of the 
corresponding eigenvalue of the Neumann matrices:
It simply moves out the spectrum; in other words,
the corresponding eigenvectors become vanishing
in this limit.  
 
Concerning some details, we would like to 
emphasize the following points:

1. The $\theta$-spectrum (noncommutativity) for 
the continuous Moyal pairs is positive and always 
bounded from above (less than $2$). In contrast, 
the noncommutativity parameters for the two 
discrete pairs have opposite sign, one of them 
is positive and bounded from below (larger than 
$2$), while the other is negative and bounded 
from above(less than $-2$). Presumably this is 
related to the fact that the two endpoints of 
the open string are "oppositely charged". 

2. In the presence of a background $B$-field, 
the Moyal pairs mix the string modes in the 
noncommutative directions parallel to the 
$B$-field. It is interesting to note that, 
for example, $x^{1}_{2n}$ is mixed with 
$p^{2}_{2n+1}$. Suppressing the mode index, 
this is the correct structure for the guiding 
center coordinates of a charged particle in 
a magnetic field. 

3. When the continuous parameter $\ka$ 
approaches to zero, only one Moyal coordinate 
for each "dimension" survives. It is either
${x}^{\mu}$ or ${y}^{\bar\alpha}$ that gives 
a commuting coordinate. This feature does 
not depend on whether the zero modes and a  
$B$-background are included or not.

4. As we mentioned above, the large-$B$ 
contraction is singular in one aspect: Namely
one of the discrete Moyal pair does not survive 
this contraction. We feel perhaps some caution 
has to be taken for some arguments that exploit
the large-$B$ limit in the open string field
theory. In other words. at finite and large $B$ 
for such arguments to work, one needs to show 
that the contributions from the dropped-out modes 
are indeed negligible.

\section*{Acknowledgments}

The authors thank Rui-Hong Yue for his 
participation in an early stage of this work 
during his short visit to Utah. We also thank
Guang-Hong Chen, Bo Feng for discussions. This research 
was supported in part by the US NSF Grant No. 
PHY-9970701.

\appendix
\section{The properties of Neumann matrices
${\cal V}^{rs,\alpha\beta}$ and 
$\MM^{rs,\alpha\beta}$}

In this appendix, we list the properties of  
Neumann matrices ${\cal V}^{rs,\alpha\beta}$ 
and $\MM^{rs,\alpha\beta}$, which are quoted 
from ref. \cite{BMS,FHM}: $(m,n>0)$ 
\ben\label{A1}
{\cal V}^{rs,\alpha\beta}_{00}
&=&G^{\alpha\beta}\delta^{rs}-\Omega b
(G^{\alpha\beta}\phi^{rs}
+\Xi\epsilon^{\alpha\beta}\chi^{rs})\,,\nonumber\\
{\cal V}^{rs,\alpha\beta}_{0n}&=&\Omega \sqrt{b}
\sum_{t=1}^{3}(G^{\alpha\beta}\phi^{rt}
+\Xi\epsilon^{\alpha\beta}\chi^{rt})V^{ts}_{0n}\,,\\ 
{\cal V}^{rs,\alpha\beta}_{mn}
&=&G^{\alpha\beta}V^{rs}_{mn}-\Omega 
\sum_{t,v=1}^{3}V^{rv}_{m0}(G^{\alpha\beta}
\phi^{vt}+\Xi\epsilon^{\alpha\beta}\chi^{vt})V^{rs}_{0n}
\ ,\nonumber
\een
where $V^{rs}_{mn}$ is Neumann matrices in the zero 
momentum sector, and $V^{rs}_{m0}$ and $V^{rs}_{0m}$ 
can be written in the following form \cite{RSZ1,Okuyama}
\ben
&&V^{rr}_{n0}=-\frac{2\sqrt{2}}{3}|v_e\rangle\ ,
\qquad\qquad V^{rr}_{0n}
=-\frac{2\sqrt{2}}{3}\langle v_e|\ ,\nonumber \\
&&V^{21}_{n0}=\frac{\sqrt2}{3}|v_e\rangle
+\frac{\sqrt6}{3}|v_o\rangle\ ,\quad
V^{12}_{0n}=\frac{\sqrt2}{3}\langle 
v_e|+\frac{\sqrt6}{3}\langle v_o|\ ,
\nonumber \\
&&V^{12}_{n0}=\frac{\sqrt2}{3}|v_e\rangle
-\frac{\sqrt6}{3}|v_o\rangle\ ,\quad
V^{21}_{0n}=\frac{\sqrt2}{3}\langle v_e|
-\frac{\sqrt6}{3}\langle v_o|\ .
\een
with $|v_{e,o}\rangle$ defined as follows: 
\be
|v_e\rangle_n=\frac{1}{\sqrt{n}}
\frac{(1+(-1)^n)}{2}A_n\ ,\quad|v_o\rangle_n
=\frac{1}{\sqrt{n}}\frac{(1-(-1)^n)}{2}A_n\ , 
\ee
where $A_n$ is the coefficients of the series 
expansion
\be
\Bigl(\frac{1+ix}{1-ix}\Bigr)^{\frac13}=
\sum_{n=\mbox{\tiny even}}A_n x^n
+i\sum_{n=\mbox{\tiny odd}}A_n x^n\ . \nonumber 
\ee
The other quantities in \refb{A1} are
\be
\Omega=\frac{2\beta}{4\pi^4 \alpha'^4 B^2 
+3\beta^2}\,,\qquad \Xi=i\frac{\pi^2\alpha'^2 B}
{\xi\beta}\,,\qquad \beta=\ln{27\over 16}+{b\over 2}\,,
\ee
and
\ben
 \chi^{rs}=\left( \begin{array}{ccc} 
0 & 1 & -1 \\ -1 & 0
& 1 \\ 1 & -1 & 0 \end{array} \right),\quad
\phi^{rs}=\left( \begin{array}{ccc}  
1 & -{1\over 2} & -{1\over 2} \\
-{1\over 2} & 1 & -{1\over 2}  \\ 
-{1\over 2} & -{1\over 2} & 1 
\end{array} \right),
\een

We have 
\ben
&& ({\cal V}^{rs,\alpha\beta})^t
={\cal V}^{sr,\beta\alpha}\ , \quad
({\MM}^{rs,\alpha\beta})^t={\MM}^{sr,\beta\alpha}\ , 
\label{A2}\\
&& C{\cal V}^{rs,\alpha\beta}
={\cal V}^{sr,\beta\alpha}C\ , \quad
C{\MM}^{rs,\alpha\beta}={\MM}^{sr,\beta\alpha}C\ ,
\label{A3}\\
&&[{\cal V}^{rs},{\cal V}^{r's'}]=
0=[{\MM}^{rs},{\MM}^{r's'}]\ .
\label{A4}
\een
And
\ben
&& \MM^{11}+\MM^{12}+\MM^{21}=
\frac{1}{\xi}\bf I \ ,\label{A5}\\
&& (\MM^{11})^2+(\MM^{12})^2+(\MM^{21})^2
=\frac{1}{\xi^2}\bf I \ ,
\label{A6}\\
&& \MM^{12}\MM^{21}=\MM^{11}(\MM^{11}
-\frac{1}{\xi}\bf I)\ ,\label{A7}
\een
where $\bf I$ stands for $\delta^{\alpha\beta}
\delta_{mn}$ and matrix multiplication is 
understood both for the indices $m,n$ and
$\alpha,\beta$. It is easy to get the 
following equations from above properties:
\ben
&& \MM^{12}+\MM^{21}
=\frac{1}{\xi}\bf I-\MM^{11} \ ,\nonumber\\
&&(\MM^{12}-\MM^{21})^2=(\frac{1}{\xi}{\bf I}
+3\MM^{11})(\frac{1}{\xi}{\bf I}-\MM^{11})\ .
\label{A8}
\een 

If we set $B=0$, the expressions and the properties 
of matrices ${\cal V}^{rs}\ ,\MM^{rs}$ will recover 
those of $V'^{rs}\ ,M'^{rs}$.

\section{The small and large $B$ limit of eigenvectors}

These two limiting case for the continuous eigenvectors 
are straightforward; in this Appendix, we will concentrate 
on the discrete spectrum.

The two discrete eigenvalues are determined by the 
following equations, respectively, with 
$x\equiv\xi \lambda$:
\be \label{B3}
2h(x)=\mp 4B\pi^2\sqrt{\frac{1-x}{1+3x}}
+b-4(\gamma+\log(4))\ ,
\ee
$x_1$ is the solution of eq. \refb{B3} with $"-"$ 
sign in it and $x_2$ $"+"$ sign. The eigenvectors 
for them are (quoted from Ref. \cite{FHM})
\ben
&&X_e(j)=i(-\frac{2\sqrt{6b}\Omega\Xi\xi}
{3(\xi\lambda_j-1+\Omega b)}\ ,
\ \frac{\xi d_{oe}}{\xi\lambda_j-M}|v_e\rangle \ ,\ 0\ ,
\ \frac{\xi d_{oo}-\xi A_{ee}(d_{oo}d_{ee}+d_{oe}^2)}
{\xi \lambda_j-M}|v_o\rangle )^t\ ,\nonumber \\[3mm]
&&X_o(j)=(0\ ,\ -\frac{\xi d_{oo}-\xi A_{ee}(d_{oo}d_{ee}+d_{oe}^2)}
{\xi \lambda_j-M}|v_o\rangle \ ,\ 
-\frac{2\sqrt{6b}\Omega\Xi\xi}
{3(\xi\lambda_j-1+\Omega b)}\ ,
\ \frac{\xi d_{oe}}{\xi\lambda_j-M}|v_e\rangle )^t \ .\nonumber 
\een
where ($\alpha'=1$)
\ben 
&&\Omega=\frac{2\beta}{4\pi^4 B^2+3\beta^2}
\ ,\quad \Xi=i\frac{\pi^2}{\xi\beta} B\ .\nonumber \\
&&d_{oo}=-\frac{2\Omega}{3\xi}
[-3+\frac{4b\Omega\Xi^2\xi^2}{\xi\lambda-1+\Omega b}]\ , 
\nonumber \\
&&d_{oe}=-\frac{4\Omega \Xi(\xi\lambda-1)}
{\sqrt 3 (\xi\lambda-1+\Omega b)}\ ,\nonumber \\
&&d_{ee}=-\frac{2\Omega(x-1)}{\xi(x-1+\Omega b)}
\ ,\nonumber \\
&&A_{ee}=-\frac{\xi}{4(x-1)}
\{9(x-1)\ln3+16\ln2+(1+3x)(h(x)+2\gamma)\}\ .
\een

\subsection{The small-$B$ limit}

Let us consider the small-$B$ limit and expand the
discrete eigenvalue to first order in $B$ 
\ben \label{B1}
x_i(B)&=&\xi \lambda_i(B)=(\xi\lambda_i)|_{B=0}+
B(\xi\lambda_i)'|_{B=0}+O(B^2)\nonumber \\
       &=&\rho_0+Bx^{(1)}_i+O(B^2)
\een

We also expand the function $h(x)$ around $\rho_0$ 
to first order in $B$,
\ben \label{B2}
h(x)&=&h(\rho_0)+(x-\rho_0)h'(\rho_0)
+{1\over 2}(x-\rho_0)^2 h''(\rho_0)+...\nonumber \\
&=&{1\over 2}b-2(\gamma+\log(4))
+Bx^{(1)}h'(\rho_0)+O(B^2)
\een 

Expanding the RHS of \refb{B3} to first order in $B$ 
and comparing it with \refb{B2}, it is easy to get 
\ben\label{B4}
h'(\rho_0)x^{(1)}=\left\{
\begin{array}{rl}
-2\pi^2\sqrt{\frac{1-\rho_0}{1+3\rho_0}} & ,
\quad\mbox{for}\,\,\lambda_1\\[3mm]
+2\pi^2\sqrt{\frac{1-\rho_0}{1+3\rho_0}} & ,
\quad\mbox{for}\,\,\lambda_2
\end{array}
\right.
\een

Now, let us consider the behavior of the eigenvectors 
in this limit. We have 
\ben 
&&\Omega={2\over{3\beta}}+O(B^2)\ , \quad
\Xi=i\frac{\pi^2}{\beta}B+O(B^3)\ ,\nonumber \\
&&d_{oo}=\frac{4}{3\beta}+O(B^2)\ ,\quad
d_{oe}=
-i\frac{8\pi^2(\rho_0-1)}{\sqrt 3 \beta(3\beta(\rho_0-1)+2b)}B+O(B^2)
\nonumber \\
&&1-d_{ee}A_{ee}=\frac{1+3\rho_0}{3\beta(\rho_0-1)+2b}
h'(\rho_0)x^{(1)}B+O(B^2)
\een
From eq. \refb{B4}, we see that the leading term of $1-d_{ee}A_{ee}$
has an opposite sign for $\lambda_1$ and $\lambda_2$. 
Comparing with the expressions of $\phi_n$ 
(eq. (6.6)(6.7) in ref. \cite{FHM1}), we get 
\ben\label{B6}
&&\lim_{B\to 0} v_{2n}^1(1)=\phi_{2n}=\lim_{B\to 0}v_{2n}^1(2)\ ,
\nonumber\\ 
&&\lim_{B\to 0} v_{2n+1}^2(1)=-i\phi_{2n+1}=
-\lim_{B\to 0}v_{2n+1}^2(2)\ .
\een 
where $n\geq 0$.

\subsection{The large-$B$ limit}

In the large-$B$ limit, $x_1$ approaches to $1$ and 
$x_2$ approaches to $0$. For $x_1$, one has
\ben
&&x_1=1-x_1^{(1)}\frac{1}{B^2}+\cdots\ ,
\nonumber \\
&&h(x_1)=h(1)+(x_1-1)h'(1)+\cdots\ .
\een
Using eq. \refb{B3}, we obtain
\be
x_1=1-\frac{b^2}{4\pi^4}\frac{1}{B^2}\ .
\ee

For $x_2\sim 0$, we have $h(x\sim0)\sim 
2\ln(-\ln(|x|)$; substituting it into eq. 
\refb{B3} we get, up to a constant factor,
\be
x_2=e^{-e^{\pi^2 B}}\ .
\ee
  
Now, let us consider the eigenvectors (here we only consider $X_e(j), j=1,2$; 
the result for $X_o(j)$ can be obtained similarly). 
For $x_1$, it is easy to get
\be
X_e(1)=(1\ ,
\ \frac{\sqrt{2}b^{\frac3 2}}{2\pi^4}
\frac{1}{B^2}\frac{1}{1-M}|v_e\rangle\ ,
\ 0\ ,\ 
i\frac{\sqrt{6b}}{3\pi^2}\frac{1}{B}
\frac{1}{1-M}|v_o\rangle)^t\ .
\ee

For $x_2\sim 0$, we need to examine the
vector $|V^{x_2}_{e,o}\rangle \equiv 
\frac{1}{x_2-M}|v_{e,o}\rangle$
carefully. First, let us calculate the 
modulus of the vector $|V^{x_2}_{o}\rangle$:
\ben
I_o(2)&\equiv&\langle v_o|(\frac{1}{x_2-M})^2
|v_o\rangle \nonumber \\
&=&\int_{-\infty}^{\infty}
\frac{d\ka}{{\cal N}(\ka)} \langle v_o|\frac{1}{x_2-M}
|\ka\rangle\langle\ka|\frac{1}{x_2-M}|v_o\rangle
\nonumber \\
&=&\int_{-\infty}^{\infty}d\ka \frac{1}{(x_2-M(\ka))^2}
\frac{1}{{\cal N}(\ka)}\langle v_o|\ka\rangle^2 
\nonumber \\
&=&\frac3 2 \int_{-\infty}^{\infty}d\ka
\frac{\sinh(\frac{\pi\ka}{2})}
{\ka(1+x_2+2x_2\cosh(\frac{\pi\ka}{2}))^2}
\nonumber \\
&=&\frac3 2 \int_{-\infty}^{\infty}dt
\frac{\sinh(t)}{t(1+x_2+2x_2\cosh(t))^2}
\een
Summing over all the residues of poles $t_n=2\pi 
i(n+\frac1 2\pm i \frac{\eta}{2\pi}),n\geq 0$ with 
$\eta=\cosh^{-1}(\frac{1+x_2}{2x_2})$ yields
\ben
I_o(2)&=&\frac{3}{2\pi^2}\frac{\eta\sinh(\eta)}
{(1-x_2)(1+3x_2)} \sum_{n=0}^{\infty}
\frac{n+\frac1 2}{[(n+\frac1 2)^2+\frac{\eta^2}
{4\pi^2}]^2} \nonumber \\
&=&\frac{3\eta}{4\pi^2 x_2}
\sqrt{(1-x_2)(1+3x_2)}S_1(\eta)\ .
\een

Similarly, we can calculate the modulus of 
the vector $|V^{x_2}_e\rangle$:
\ben
I_e(2)&\equiv&\langle v_e|(\frac{1}{x_2-M})^2
|v_e\rangle 
\nonumber \\
&=&\frac1 2 \int_{-\infty}^{\infty}dt
\frac{(\cosh(t)-1)^2}
{t\sinh(t)(1+x_2+2x_2\cosh(t))^2}
\een
Summing over all the residues of the poles
$t_n=2\pi i (n+\frac12)\ ,\quad 2\pi 
i(n+\frac1 2\pm i \frac{\eta}{2\pi})\ ,\ n\geq 0$ 
yields
\ben
I_e(2)&=&\frac{1}{(1-x_2)^2}\{
-\frac{\eta^2}{2\pi^2}\sum_{n=0}^{\infty}
\frac{1}{(n+\frac12)[(n+\frac12)^2
+\frac{\eta^2}{4\pi^2}]}
\nonumber \\
&&+\frac{\eta}{4\pi^2 x_2}\sqrt{(1-x_2)(1+3x_2)}
\sum_{n=0}^{\infty}\frac{n+\frac12}
{[(n+\frac12)^2 +\frac{\eta^2}{4\pi^2}]^2} \}
\nonumber \\
&=&\frac{1}{(1-x_2)^2}
\{\frac{\eta}{4\pi^2 x_2}\sqrt{(1-x_2)(1+3x_2)}
S_1(\eta)-\frac{\eta^2}{2\pi^2}S_2(\eta)
\} 
\een

Let us estimate the summation $S_1(\eta)$ 
as follows:
\ben
\frac{\frac1 2}{(\frac1 4
+\frac{\eta^2}{4\pi^2})^2} <& S_1(\eta) &<
\frac{\frac1 2}{(\frac1 4
+\frac{\eta^2}{4\pi^2})^2}+\int_0^\infty dx
\frac{x+\frac1 2}{[(x+\frac1 2)^2
+\frac{\eta^2}{4\pi^2}]^2} \nonumber \\
\frac{8\pi^4}{(\pi^2+\eta^2)^2} <& S_1(\eta) &<
\frac{8\pi^4}{(\pi^2+\eta^2)^2}
+\frac{2\pi^2}{\pi^2+\eta^2}\ .
\een
and the summation $S_2(\eta)$: 
\ben
\frac{1}{\frac12 (\frac1 4+\frac{\eta^2}{4\pi^2})}
<& S_2(\eta) &< \frac{1}{\frac12 (\frac1 4
+\frac{\eta^2}{4\pi^2})}+\int_0^\infty dx
\frac{1}{(x+\frac1 2)[(x+\frac1 2)^2
+\frac{\eta^2}{4\pi^2}]} \nonumber\\
\frac{8\pi^2}{(\pi^2+\eta^2)} <& S_2(\eta) &<
\frac{8\pi^2}{(\pi^2+\eta^2)}+
\frac{4\pi^2}{\eta^2}\ln(1+\frac{\eta^2}{\pi^2})\ .
\een

So, for $x_2=e^{-e^{\pi^2 B}}\sim 0$, we see that
\be
I_o(2)\sim e^{e^{\pi^2 B}}\ ,\quad I_e(2)
\sim e^{e^{\pi^2 B}}
\ee

The normalized eigenvector for $x_2$ is then
\be
X_e(2)\sim 
(0\ ,\ 2\sqrt3 \frac{e^{-\frac12 e^{\pi^2 B}}}
{x_2-M}|v_e\rangle\ ,\ 0\ ,\ i\frac{e^{-\frac12 
e^{\pi^2 B}}}{x_2-M}|v_o\rangle)^t\ .
\ee
Expanding it in the $|\ka\rangle$ basis, we have 
\ben
\frac{e^{-\frac12 e^{\pi^2 B}}}{x_2-M}|v_o\rangle 
&=& e^{-\frac12 e^{\pi^2 B}}
\int_{-\infty}^{\infty}\frac{d\ka}{{\cal N}(\ka)}
|\ka\rangle\langle \ka|\frac{1}{x_2-M}|v_o\rangle 
\nonumber \\
&=& e^{-\frac12 e^{\pi^2 B}}\int_{-\infty}^{\infty}
{d\ka} \sqrt{\frac{3\sinh(\frac{\pi\ka}{2})}{2\ka}}
\frac{1}{1+x_2+2x_2\cosh(\frac{\pi\ka}{2})}|\td\ka
\rangle\ ,\nonumber \\
\frac{e^{-\frac12 e^{\pi^2 B}}}{x_2-M}|v_e\rangle 
&=& e^{-\frac12 e^{\pi^2 B}}
\int_{-\infty}^{\infty}\frac{d\ka}{{\cal N}(\ka)}
|\ka\rangle\langle \ka|\frac{1}{x_2-M}|v_e\rangle 
\nonumber \\
&=& e^{-\frac12 e^{\pi^2 B}}\int_{-\infty}^{\infty}
{d\ka}\tanh(\frac{\pi\ka}{4})
\sqrt{\frac{\sinh(\frac{\pi\ka}{2})}{2\ka}}
\frac{1}{1+x_2+2x_2\cosh(\frac{\pi\ka}{2})}|\td\ka
\rangle\ .
\een
where $|\td \ka\rangle$ is the normalized eigenvector 
of the Neumann matrix $M$. We see that the larger the 
$\ka$, the bigger the coefficient in front of the 
basis vector $|\td\ka\rangle$. However, for fixed 
$\ka$, the coefficient in front of the basis vector 
$|\td\ka\rangle$ goes to zero as $B\to\infty$. So we 
conclude that as $B\to\infty$, the contribution can 
only come from $\ka=\infty$, which is however known 
to be not in the spectrum of $M$ \cite{RSZ}. In other words,
since the Neumann matrix $M$ has no nontrivial 
eigenvector with eigenvalue $\ka=\infty$, the 
normalized vectors $X_{e,o}(2)$ will move out the string 
Hilbert space as $B\to \infty$. (The non-normalized
vector will have all components vanishing in the 
limit.) 

Thus, only the eigenvalue $x_1$ survives the large-$B$ 
limit, while the behavior of the eigenvalue $x_2$ is 
singular: It moves out the spectrum.

\section{The discrete eigenvectors of 
$\MM^{12}(\MM^{21})$}

Let us prove that
\ben \label{C1}
\MM^{rs}X_{\pm}(j)&=&\lambda^{rs}_{j,\pm}X_{\pm}(j)\ ,
\een
where $X_{\pm}(j)=X_e(j)\pm X_o(j),\ j=1,2$.

First, from eq. (A.7), we know 
\ben \label{C2}
\MM^{11}(\MM^{12}-\MM^{21})X_{e,o}(j)&=&
(\MM^{12}-\MM^{21})\MM^{11}X_{e,o}(j)\nonumber \\
&=&\lambda^{11}_j (\MM^{12}-\MM^{21})X_{e,o}(j)\,.
\een
So, $(\MM^{12}-\MM^{21})X_{e,o}(j)$ should be a
linear combination of the eigenvectors of 
$\MM^{11}$, i.e.,
\ben \label{C3}
(\MM^{12}-\MM^{21})X_e(j)
=a_{1j}X_e(j)+b_{1j}X_o(j) \,,\nonumber \\
(\MM^{12}-\MM^{21})X_o(j)
=a_{2j}X_e(j)+b_{2j}X_o(j) \,.
\een

Using $C\MM^{rs,\alpha\beta}C=\MM^{sr,\beta\alpha}$ 
and expanding \refb{C3} explicitly, we can get 
immediately
\be
a_{1j}=b_{2j}=0\,,\;\;\;a_{2j}=b_{1j}=\eta_j\,.
\ee
Thus, the equations \refb{C3} now read
\ben \label{C4}
(\MM^{12}-\MM^{21})X_e(j)=\eta_j X_o(j) \,,\nonumber \\
(\MM^{12}-\MM^{21})X_o(j)=\eta_j X_e(j) \,.
\een
and we can further get
\be \label{C5}
(\MM^{12}-\MM^{21})^2 X_{e,o}(j)=\eta_j^2 X_{e,o}(j) \,.
\ee

On the other hand, we know that $(\MM^{12}-\MM^{21})^2
=(\frac{1}{\xi}+3\MM^{11})(\frac{1}{\xi}-\MM^{11})$,
so 
\be
\eta_j^2=(\frac{1}{\xi}+3\lambda_j)
(\frac{1}{\xi}-\lambda_j) \,.
\ee 
Thus,
\ben
\MM^{12}X_+(j)&&=
[\frac 12 (\MM^{12}-\MM^{21})+
\frac 12 (\MM^{12}+\MM^{21})][X_e(j)+ X_o(j)] 
\nonumber \\
&&=[\frac 12 \eta_j+\frac 12 (\frac{1}
{\xi}-\lambda_j)][X_e(j)+X_o(j)] \nonumber \\
&&=\lambda^{12}_{j,+}X_+ (j)\ .
\een
Similarly, we get
\ben
\MM^{12}X_- (j)=\lambda^{12}_{j,-}X_- (j)\ ,\\
\MM^{21}X_+ (j)=\lambda^{21}_{j,+}X_+ (j)\ ,\\
\MM^{21}X_- (j)=\lambda^{21}_{j,-}X_- (j)\ ,
\een
where
\ben\label{evdB12}
\lambda^{12}_{j,\pm}=\pm\frac 12 \eta_j
+\frac 12(\frac{1}{\xi}-\lambda_j)
\ ,\\
\lambda^{21}_{j,\pm}=\mp\frac 12 \eta_j
+\frac 12 (\frac{1}{\xi}-\lambda_j)
\ .
\een

In order to determine the sign in front of 
$\sqrt{\eta_j^2}$, first we take the limit 
$B\to 0$ in eqs. \refb{C4}:
\ben
&&\sum_{n=0}^{\infty}M^{12}_{2m,2n+1}
\left[iv^2_{2n+1}(j)\right]_{B=0}
=\left[\eta_j v^1_{2m}(j)\right]_{B=0} \ ,
\nonumber\\[2mm]
&&\sum_{n=0}^{\infty}M^{12}_{2m+1,2n}[v^1_{2n}(j)]_{B=0}
=\left[\eta_j \left(iv^2_{2m+1}(j)\right)\right]_{B=0} \ .\nonumber
\een
Recall eqs. \refb{B6}, we obtain 
\be
\eta_1|_{B=0}=-\eta_2|_{B=0} \ . \nonumber
\ee
Then taking the limit $B\to 0$ in eq. \refb{evdB12} 
and comparing it with eq. \refb{evd012}, we get
\be
\eta_1=\frac 12
\sqrt{(\frac{1}{\xi}+3\lambda_1)
(\frac{1}{\xi}-\lambda_1)}\ .
\ee

\end{document}